# Secure Communication over Fading Channels [1] [2]

Yingbin Liang, H. Vincent Poor and Shlomo Shamai (Shitz) [3]

**Abstract**

The fading broadcast channel with confidential messages (BCC) is investigated, where a source node has common information for two receivers (receivers 1 and 2), and has confidential information intended only for receiver 1. The confidential information needs to be kept as secret as possible from receiver 2. The broadcast channel from the source node to receivers 1 and 2 is corrupted by multiplicative fading gain coefficients in addition to additive Gaussian noise terms. The channel state information (CSI) is assumed to be known at both the transmitter and the receivers. The parallel BCC with independent subchannels is first studied, which serves as an information-theoretic model for the fading BCC. The secrecy capacity region of the parallel BCC is established. This result is then specialized to give the secrecy capacity region of the parallel BCC with degraded subchannels. The secrecy capacity region is then established for the parallel Gaussian BCC, and the optimal source power allocations that achieve the boundary of the secrecy capacity region are derived. In particular, the secrecy capacity region is established for the basic Gaussian BCC. The secrecy capacity results are then applied to study the fading BCC. The ergodic performance is first studied. The ergodic secrecy capacity region and the optimal power allocations that achieve the boundary of this region are derived. The outage performance is then studied, where a long term power constraint is assumed. The power allocation is derived that minimizes the outage probability where either the target rate of the common message or the target rate of the confidential message is not achieved. The power allocation is also derive that minimizes the outage probability where the target rate of the confidential message is not achieved subject to the constraint that the target rate of the common message must be achieved for all channel states.

---

[1]The material in this paper was presented in part at the 44th Annual Allerton Conference on Communication, Control, and Computing, Monticello, IL, Sept. 2006.

[2]The research was supported by the National Science Foundation under Grant ANI-03-38807 and by National Science Foundation.

[3]Yingbin Liang and H. Vincent Poor are with the Department of Electrical Engineering, Princeton University, Engineering Quadrangle, Olden Street, Princeton, NJ 08544; e-mail: {yingbinl,poor}@princeton.edu; Shlomo Shamai (Shitz) is with the Department of Electrical Engineering, Technion-Israel Institute of Technology, Technion City, Haifa 32000, Israel; email: sshlomo@ee.technion.ac.il

# 1 Introduction

Wireless communication has a broadcast nature, where security issues are captured by a basic wire-tap channel introduced by Wyner in [1]. In this model, a source node wishes to transmit confidential information to a destination node and wishes to keep a wire-tapper as ignorant of this information as possible. The performance measure of interest is the secrecy capacity, which is the largest reliable communication rate from the source node to the destination node with the wire-tapper obtaining no information. The secrecy capacity was given in [1] for the discrete memoryless wire-tap channel and in [2] for the Gaussian wire-tap channel. The wire-tap channel was considered recently for the fading and multiple antenna channels in [3, 4]. A more general model of the wire-tap channel was studied by Csiszár and Körner in [5], where the source node also has a common message for both receivers in addition to the confidential message for only one receiver. This channel is regarded as the broadcast channel with confidential messages (BCC). The capacity-equivocation region and the secrecy capacity region of the BCC were characterized in [5]. The BCC was further studied recently in [6], where the source node transmits two confidential message sets for two receivers, respectively, in addition to the common message for both receivers.

In this paper, we investigate the fading BCC, which is based on the BCC studied in [5] with the channels from the source node to receivers 1 and 2 corrupted by multiplicative fading gain coefficients in addition to additive Gaussian noise terms. The fading BCC model captures the basic time-varying property of wireless channels, and hence understanding this channel plays an important role in solving security issues in wireless applications. For the fading BCC, we assume that the fading gain coefficients are stationary and ergodic over time. We further assume that the channel state information (CSI) is known at both the transmitter and the receivers. The CSI at the source node can be realized by a reliable feedback from the two receivers, who are supposed to receive information from the source node.

The fading BCC we study in this paper relates to or generalizes a few channels that have been previously studied in the literature. Compared to the fading broadcast channel that was studied in [7–11], the fading BCC requires an additional secrecy constraint that the confidential information for one receiver must be perfectly secret from the other receiver.

Compared to the fading wire-tap channel studied in [12] (the conference version of this paper), [13] and [14] (full CSI case), the fading BCC we study in this paper assumes that the source node has a common message for both receivers in addition to the confidential message for receiver 1. Hence the fading BCC includes the fading wire-tap channel as a special case. The fading BCC also includes the parallel Gaussian wire-tap channel studied in [15] (the case where wire-tappers cooperate) as a special case for the same reason as above and also because a power constraint is assumed for each subchannel in [15].

Before studying the fading BCC, we first study a more general model of the parallel BCC with $L$ independent subchannels, where the source node communicates with receivers 1 and 2 over $L$ parallel links. This model serves as a general information-theoretic model that includes the fading BCC as a special case. We establish the secrecy capacity region of the parallel BCC. In particular, we provide a converse proof to show that independent input distribution for each subchannel is optimal to achieve the secrecy capacity region. This fact does not follow directly from the single letter characterization of the secrecy capacity region of the BCC given in [5]. The secrecy capacity region of the parallel BCC further gives the secrecy capacity region of the parallel BCC with degraded subchannels.

We further study the parallel Gaussian BCC, which is an example parallel BCC with degraded subchannels. We show that the secrecy capacity region of the parallel Gaussian BCC is a union over the rate regions achieved by all source power allocations (among the parallel subchannels). Moreover, we derive the optimal power allocations that achieve the boundary of the secrecy capacity region and hence completely characterize this region. The secrecy capacity region of the parallel Gaussian BCC also establishes the secrecy capacity region for the basic Gaussian BCC. This result complements the secrecy capacity region of the discrete memoryless BCC given by Csiszár and Körner in [5].

We then apply our results to investigate the fading BCC. We first study the ergodic performance, where no delay constraints on message transmission are assumed and the secrecy capacity region is averaged over all channel states. Now the fading BCC can be viewed as the parallel Gaussian BCC with each fading state corresponding to one subchannel. Thus, the secrecy capacity region of the parallel Gaussian BCC applies to the fading BCC. In particular, since the source node knows the CSI, it can dynamically change its transmission power with channel state realization to achieve the best performance. We obtain the optimal power allocations that achieve the boundary of the secrecy capacity region for the fading BCC.

We further study the outage performance of the fading BCC, where messages must be transmitted over a certain time (one block) to satisfy the delay constraint. We adopts the block fading model, where the fading coefficients remain constant over one block and change to another realization in the next block. The block length is assumed to be large enough to guarantee decoding in one block. We assume the power constraint at the source node applies over many blocks (i.e., it is a long term power constraint as in [16]). As in the analysis of the ergodic performance, we assume that the CSI is known both at the transmitter and at the receivers, and hence the source node can allocate its transmission power to achieve the best outage performance. We first obtain the power allocation that minimizes the outage probability where either the target rate of the common message or the target rate of the confidential message is not achieved. We then obtain the power allocation that minimizes the outage probability where the target rate of the confidential message is not achieved subject to the constraint that the target rate of the common message must be achieved for all channel states.

In this paper, we use $X_{[1,L]}$ to indicate a group of variables $(X_1, X_2, \ldots, X_L)$, and use $X^n_{[1,L]}$ to indicate a group of vectors $(X^n_1, X^n_2, \ldots, X^n_L)$, where $X^n_l$ indicates the vector $(X_{l1}, X_{l2}, \ldots, X_{ln})$. Throughout the paper, the logarithmic function is to the base 2.

The paper is organized as follows. We first study the parallel BCC with independent subchannels, and its special case of the parallel BCC with degraded subchannels. We next study the parallel Gaussian BCC. We then study the ergodic and outage performances of the fading BCC and demonstrate our results with numerical examples. We conclude the paper with a few remarks.

We note that the wire-tap channel has also been studied in the work [17–27] and references therein. The topic of common randomness and secret key capacity in communication systems has been studied in [28–30]. Such communication system may be viewed as a wire-tap channel with side information (which might be common randomness, i.e., a key at a certain rate). Other related work on this topic can be found in [31–39] and references therein. We also remark that the secrecy rate/capacity has also been studied for the multiple access channel [40–43] the relay channel in [44], and the interference channel in [6].

# 2 Parallel BCCs

## 2.1 Channel Model

We consider the parallel BCC with $L$ independent subchannels (see Fig. 1), where there are one source node and two receivers. Each subchannel is assumed to be a general broadcast channel from the source node to the two receivers. As in the BCC, the source node wants to transmit common information to both receivers and confidential information to receiver 1. Moreover, the source node wishes to keep the confidential information to be as secret as possible from receiver 2.

More formally, the parallel BCC consists of $L$ finite input alphabets $\mathcal{X}_{[1,L]}$, and $2L$ finite output alphabets $\mathcal{Y}_{[1,L]}$ and $\mathcal{Z}_{[1,L]}$. The transition probability distribution is given by

$$p(y_{[1,L]}, z_{[1,L]}|x_{[1,L]}) = \prod_{l=1}^{L} p_l(y_l, z_l|x_l) \tag{1}$$

where $x_l \in \mathcal{X}_l$, $y_l \in \mathcal{Y}_l$, and $z_l \in \mathcal{Z}_l$ for $l = 1, \ldots, L$.

If the parallel BCC has only one subchannel, i.e., $L = 1$, this channel becomes the BCC studied in [5]. Moreover, each subchannel is assumed to be a general broadcast channel as in [5] and is not necessarily degraded as assumed in [1].

A $\left(2^{nR_0}, 2^{nR_1}, n\right)$ code consists of the following:

- Two message sets: $\mathcal{W}_0 = \{1, 2, \ldots, 2^{nR_0}\}$ and $\mathcal{W}_1 = \{1, 2, \ldots, 2^{nR_0}\}$ with the messages $W_0$ and $W_1$ uniformly distributed over the sets $\mathcal{W}_0$ and $\mathcal{W}_1$, respectively;
- One (stochastic) encoder at the source node that maps each message pair $(w_0, w_1) \in (\mathcal{W}_0, \mathcal{W}_1)$ to a codeword $x^n_{[1,L]}$;
- Two decoders: one at receiver 1 that maps a received sequence $y^n_{[1,L]}$ to a message pair $(\hat{w}_0^{(1)}, \hat{w}_1) \in (\mathcal{W}_0, \mathcal{W}_1)$; the other at receiver 2 that maps a received sequence $z^n_{[1,L]}$ to a message $\hat{w}_0^{(2)} \in \mathcal{W}_0$.

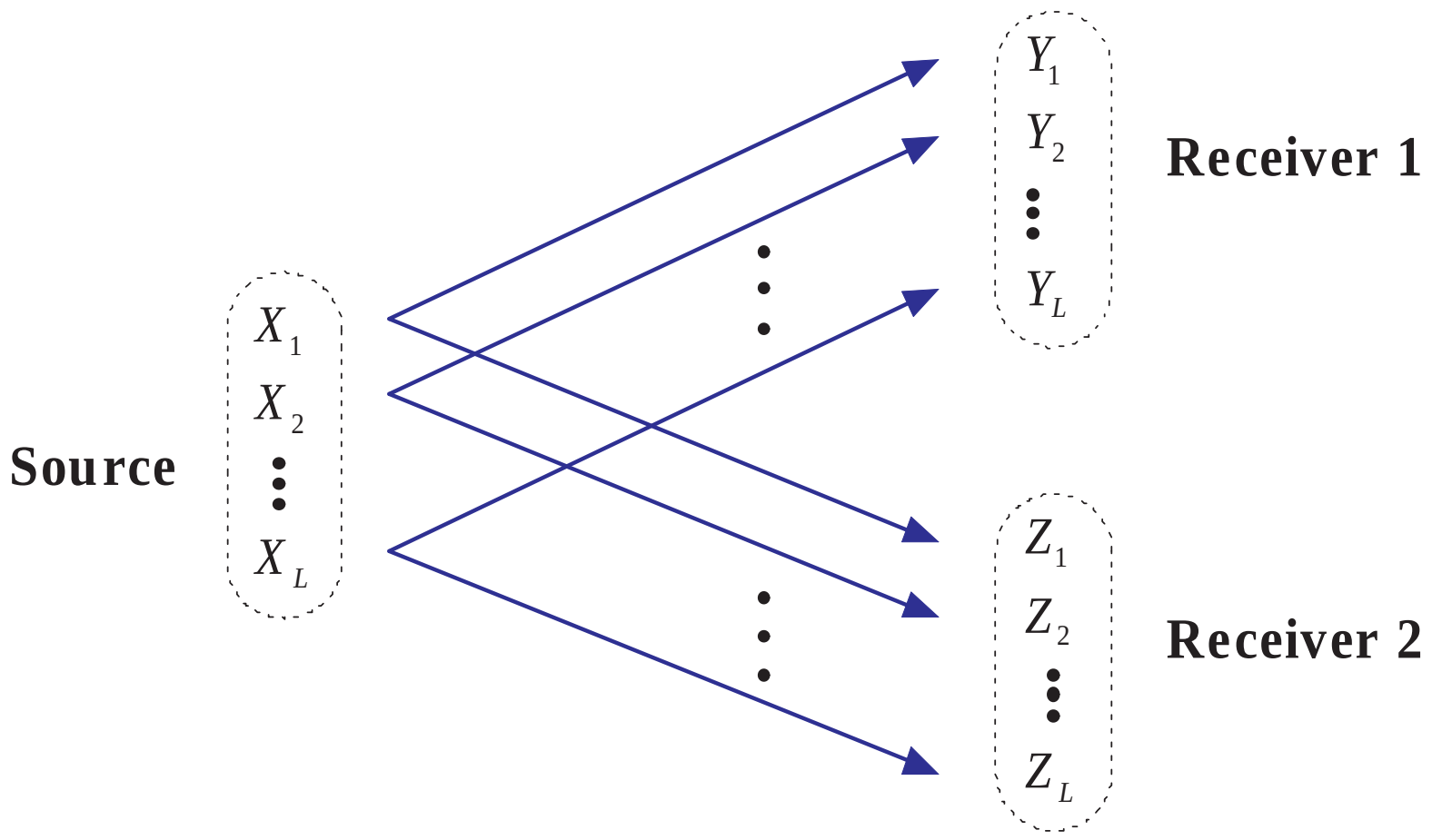

Figure 1: Parallel BCC

The secrecy level of the confidential message $W_1$ achieved at receiver 2 is measured by the *equivocation rate* defined as follows:

$$\frac{1}{n}H\left(W_1 \middle| Z_{[1,L]}^n\right). \tag{2}$$

The higher the equivocation rate, the less information that receiver 2 obtains about the confidential message $W_1$.

The probability of error when the message pair $(w_0, w_1)$ is sent is defined as

$$P_e(w_0, w_1) = Pr\left\{(\hat{w}_0^{(1)}, \hat{w}_1) \neq (w_0, w_1) \text{ or } \hat{w}_0^{(2)} \neq w_0\right\} \tag{3}$$

and the average probability of error is

$$P_e = \frac{1}{2^{nR_0}2^{nR_1}} \sum_{w_0=1}^{2^{nR_0}} \sum_{w_1=1}^{2^{nR_1}} P_e(w_0, w_1). \tag{4}$$

A rate-equivocation triple $(R_0, R_1, R_e)$ is *achievable* if there exists a sequence of $\left(2^{nR_0}, 2^{nR_1}, n\right)$ codes with the average probability of error $P_e \to 0$ as $n$ goes to infinity and with the equivocation rate $R_e$ satisfying

$$R_e \leq \lim_{n\to\infty} \frac{1}{n}H\left(W_1 \middle| Z_{[1,L]}^n\right). \tag{5}$$

In this paper, we focus on the case in which perfect secrecy is achieved, i.e., receiver 2 does not obtain any information about the message $W_1$. This happens if $R_e = R_1$. The *secrecy* *capacity region* $\mathcal{C}_s$ is defined to be the set that includes all $(R_0, R_1)$ such that $(R_0, R_1, R_e = R_1)$ is achievable, i.e.,

$$\mathcal{C}_s = \Big\{(R_0, R_1) : (R_0, R_1, R_e = R_1) \text{ is achievable}\Big\}. \tag{6}$$

## 2.2 Secrecy Capacity Region of Parallel BCCs

For the parallel BCC, we obtain the following secrecy capacity region.

**Theorem 1.** *The secrecy capacity region of the parallel BCC is given by*

$$\mathcal{C}_s = \bigcup_{\prod_l \left[p(q_l)p(u_l|q_l)p(x_l|u_l)p(y_l,z_l|x_l)\right]} \left\{ \begin{array}{l} (R_0, R_1) : \\ R_0 \le \min\left\{\sum_{l=1}^{L} I(Q_l; Y_l)\ ,\ \sum_{l=1}^{L} I(Q_l; Z_l)\right\} \\ R_1 \le \sum_{l=1}^{L} \left[I(U_l; Y_l|Q_l) - I(U_l; Z_l|Q_l)\right] \end{array} \right\} \tag{7}$$

*where* $Q_l$ *can be chosen as a deterministic function of* $U_l$ *for* $l = 1, \ldots, L$.

If the source node transmits only confidential information to receiver 1, i.e., $R_0 = 0$, the parallel BCC becomes the parallel wire-tap channel. The secrecy capacity of this channel follows from Theorem 1 by noticing that $I(U_l; Y_l|Q_l) - I(U_l; Z_l|Q_l)$ is maximized by a constant $Q_l$.

**Corollary 1.** *The secrecy capacity of the parallel wire-tap channel is*

$$C_s = \sum_{l=1}^{L} C_s^{(l)} \tag{8}$$

*where* $C_s^l$ *is the secrecy capacity of subchannel* $l$ *and is given by*

$$C_s^{(l)} = \max \left[I(U_l; Y_l) - I(U_l; Z_l)\right] \tag{9}$$

*The maximum in the preceding equation is over the distributions* $p(u_l, x_l)p(y_l, z_l|x_l)$*, which satisfies the Markov chain condition* $U_l \to X_l \to (Y_l, Z_l)$.

**Remark 1.** *Theorem 1 shows an important fact that independent input distribution for each subchannel is optimal. This fact does not follow directly from the single letter result on the secrecy capacity of the BCC given in [5], although the parallel BCC can be viewed as a special case of the BCC. Hence a converse proof is needed, which is provided in Appendix A.*

We note that the secrecy capacity region $\mathcal{C}_s^{(l)}$ of subchannel $l$ is given in [5, Corollary 1], i.e. ,

$$\mathcal{C}_s^{(l)} = \bigcup_{p(q_l)p(u_l|q_l)p(x_l|u_l)p(y_l,z_l|x_l)} \left\{ \begin{array}{l} (R_0, R_1): \\ R_0 \leq \min\{I(Q_l;Y_l)\ ,\ I(Q_l;Z_l)\} \\ R_1 \leq I(U_l;Y_l|Q_l) - I(U_l;Z_l|Q_l) \end{array} \right\}. \tag{10}$$

We now define the sum of the secrecy capacity regions of the subchannels to be

$$\mathcal{C}_{sum} := \left\{ \begin{array}{l} (R_0, R_1): \\ R_0 = \sum_l R_{l0}, \quad R_1 = \sum_l R_{l1}, \\ \text{ with } (R_{l0}, R_{l1}) \in C_s^{(l)} \text{ for } l = 1, \ldots, L. \end{array} \right\} \tag{11}$$

If we compare $\mathcal{C}_s$ with $\mathcal{C}_{sum}$, we make the following remark.

**Remark 2.** *The secrecy capacity region $\mathcal{C}_s$ in Theorem 1 may be larger than the sum $\mathcal{C}_{sum}$ of the secrecy capacity regions of the subchannels. Hence the secrecy capacity region of the parallel BCC is achieved by coding over all parallel subchannels.*

This observation was also made in [45] for the broadcast channel with common messages. This fact follows because the common rate has the following property

$$\begin{aligned} R_0 &= \min\left\{\sum_{l=1}^{L} I(Q_l;Y_l)\ ,\ \sum_{l=1}^{L} I(Q_l;Z_l)\right\} \\ &\geq \sum_{l=1}^{L} \min\{I(Q_l;Y_l)\ ,\ I(Q_l;Z_l)\} = \sum_{l=1}^{L} R_{l0} \end{aligned} \tag{12}$$

It can also be seen from the following simple example. For simplicity, we consider the case in which the source node has only the common message for both receivers. We further assume $L = 2$, and each subchannel is a deterministic broadcast channel for $l = 1, 2$ (see Fig. 2). For subchannel 1, the link capacities to receivers 1 and 2 are $C_{11} = 3$ and $C_{12} = 4$, respectively. For subchannel 2, the link capacities to receivers 1 and 2 are $C_{21} = 7$ and $C_{22} = 5$, respectively. The capacity of this parallel channel is given by

$$C = \min\{C_{11} + C_{21}, C_{12} + C_{22}\} = \min\{3 + 7, 4 + 5\} = 9. \tag{13}$$

However, the sum of the capacities of the two subchannels is

$$\sum_{l=1}^{2} \min\{C_{l1}, C_{l2}\} = \min\{3, 4\} + \min\{7, 5\} = 8 \tag{14}$$

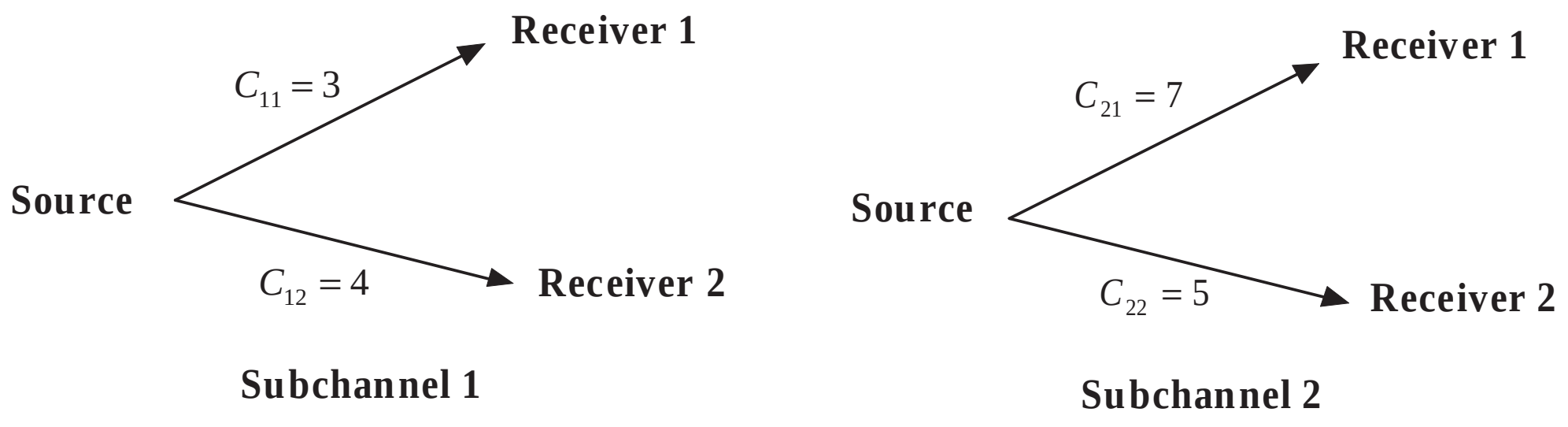

Figure 2: A parallel BCC example

which is clearly smaller than the capacity given in (13).

Similar to Theorem 14.6.1 in [46], we obtain the following lemma for the BCC studied in [5], which also applies to the parallel BCC we study in this paper.

**Lemma 1.** *The secrecy capacity region of the BCC studied in [5] depends only on the marginal transition probability distributions $p(y|x)$ of the channel from the source node to receiver 1 and $p(z|x)$ of the channel from the source node to receiver 2.*

*Proof.* The proof follows from the reasoning in [46, p. 454, Prob. 10] and the fact that the equivocation rate $\frac{1}{n}H\left(W_1 \middle| Z^n_{[1,L]}\right)$ depends only on the marginal distribution of $p(z|x)$. $\square$

One application of Lemma 1 is to obtain the following generalization of the result in [2] for the Gaussian wire-tap channel, which is a special case of the BCC.

**Corollary 2.** *The secrecy capacity of the Gaussian wire-tap channel given in [2, Theorem 1] holds for the case, where the noise variables at the destination node and the wire-tapper have a general correlation structure.*

Lemma 1 will be useful in establishing the secrecy capacity region of the parallel Gaussian BCC in Section 3.

## 2.3 Parallel BCCs with Degraded Subchannels

In the following, we study a special class of parallel BCCs, which specializes to the parallel Gaussian BCC that is considered in Section 3, and is hence of particular interest.

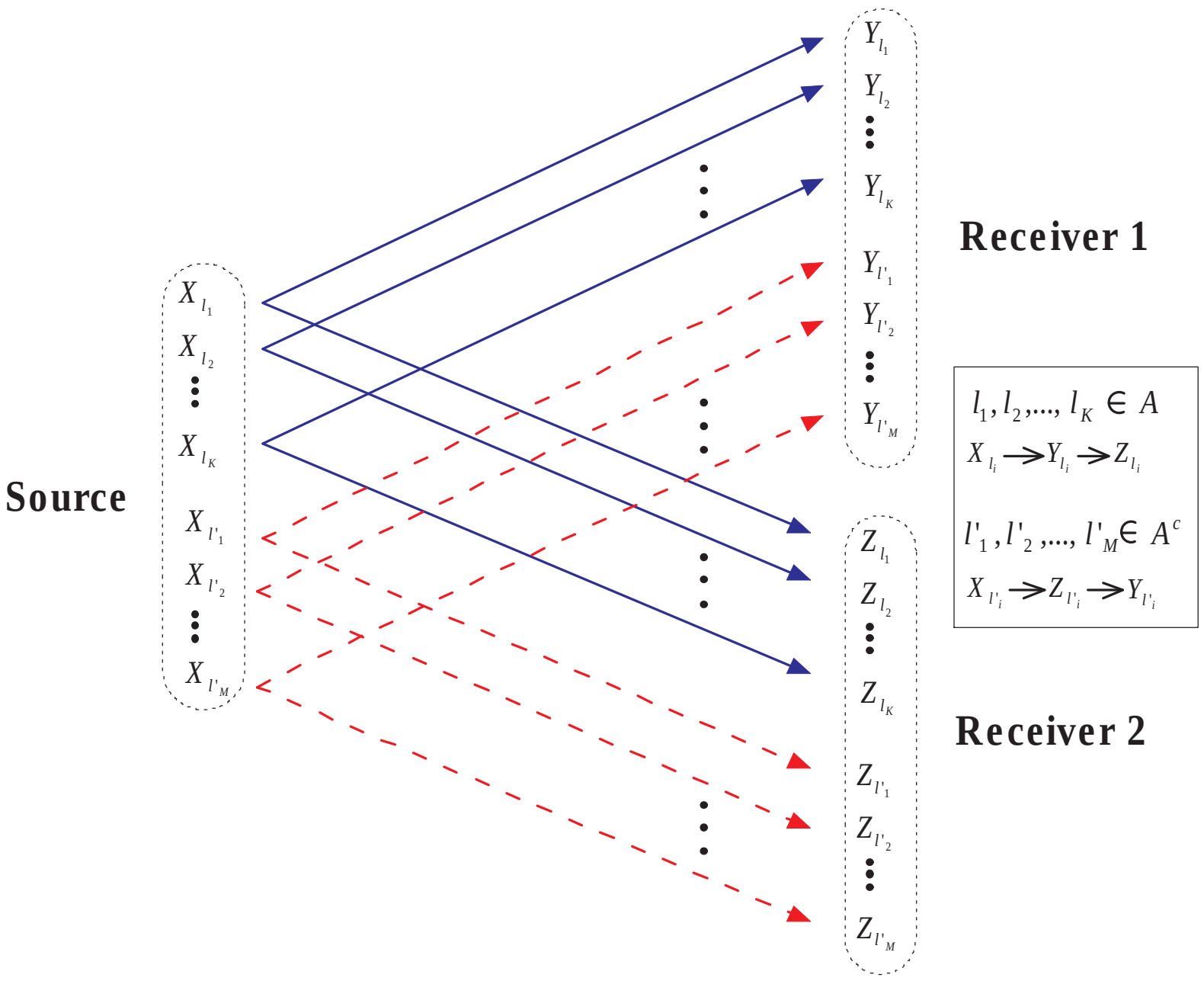

Figure 3: Parallel BCC with degraded subchannels

We consider the parallel BCC with degraded subchannels (see Fig. 3), where each subchannel is either degraded such that the output at receiver 2 is a degraded version of the output at receiver 1, or degraded such that the output at receiver 1 is a degraded version of the output at receiver 2. Note that although each subchannel is degraded, the entire channel may not be degraded because the subchannels may not be degraded in the same fashion.

We define $A$ to be the index set that includes all indices of subchannels, where the output at receiver 2 is a degraded version of the output at receiver 1, i.e.,

$$p_l(y_l, z_l|x_l) = p_l(y_l|x_l)p_l(z_l|y_l) \qquad \text{for} \quad l \in A. \tag{15}$$

Hence the Markov chain condition $X_l \rightarrow Y_l \rightarrow Z_l$ is satisfied for $l \in A$. We define $A^c$ to be the complement of the set $A$, and $A^c$ includes all indices of subchannels, where the output at receiver 1 is a degraded version of the output at receiver 2, i.e.,

$$p_l(y_l, z_l|x_l) = p_l(z_l|x_l)p_l(y_l|z_l) \qquad \text{for} \quad l \in A^c. \tag{16}$$

Hence the Markov chain condition $X_l \rightarrow Z_l \rightarrow Y_l$ is satisfied for $l \in A^c$. The channel transition probability distribution is given by

$$p(y_{[1,L]}, z_{[1,L]}|x_{[1,L]}) = \prod_{l \in A} \Big[p_l(y_l|x_l)p_l(z_l|y_l)\Big] \prod_{l \in A^c} \Big[p_l(z_l|x_l)p_l(y_l|z_l)\Big]. \tag{17}$$

For the parallel BCC with degraded subchannels, we apply Theorem 1 and obtain the following secrecy capacity region.

**Corollary 3.** *The secrecy capacity region of the parallel BCC with degraded subchannels is*

$$\mathcal{C}_s = \bigcup_{\substack{\prod_{l\in A}[p(q_l)p(u_l|q_l)p(x_l|u_l)p(y_l|x_l)p(z_l|y_l)] \\ \cdot \prod_{l\in A^c}[p(q_l)p(u_l|q_l)p(x_l|u_l)p(z_l|x_l)p(y_l|z_l)]}} \left\{ \begin{array}{l} (R_0, R_1): \\ R_0 \leq \min\left\{ \sum_{l\in A} I(Q_l;Y_l) + \sum_{l\in A^c} I(X_l;Y_l), \sum_{l\in A} I(Q_l;Z_l) + \sum_{l\in A^c} I(X_l;Z_l) \right\}, \\ R_1 \leq \sum_{l\in A} \Big[ I(X_l;Y_l|Q_l) - I(X_l;Z_l|Q_l) \Big] \end{array} \right\}. \tag{18}$$

**Remark 3.** *It can be seen that the common message $W_0$ is sent over all subchannels, and the private message $W_1$ for receiver 1 is sent only over the subchannels where the output at receiver 2 is a degraded version of the output at receiver 1, i.e., $l \in A$. Furthermore, over these subchannels, the messages $W_0$ and $W_1$ are sent by using the superposition encoding scheme.*

*Proof.* The achievability follows from Theorem 1 by setting $U_l = X_l$ for $l \in A$ and setting $Q_l = U_l = X_l$ for $l \in A^c$.

To show the converse, we first note that for $l \in A^c$,

$$\begin{aligned} I(Q_l;Y_l) &\leq I(X_l;Y_l) \\ I(Q_l;Z_l) &\leq I(X_l;Z_l) \end{aligned} \tag{19}$$

which follow from the Markov condition $Q_l \to X_l \to (Y_l Z_l)$. We apply the bounds in (19) to the bound on $R_0$ given in (7) and obtain the bound on $R_0$ given in (18).

For $l \in A^c$, we also obtain

$$\begin{aligned} & I(U_l;Y_l|Q_l) - I(U_l;Z_l|Q_l) \\ & \quad \leq I(U_l;Y_l Z_l|Q_l) - I(U_l;Z_l|Q_l) \\ & \quad = I(U_l;Z_l|Q_l) + I(U_l;Y_l|Q_l Z_l) - I(U_l;Z_l|Q_l) \\ & \quad = 0 \end{aligned} \tag{20}$$

where the last equality follows because $I(U_l;Y_l|Q_lZ_l) = 0$ due to the degraded condition (16).

For $l \in A$, we obtain the following bound

$$
\begin{aligned}
&I(U_l;Y_l|Q_l) - I(U_l;Z_l|Q_l) \\
&\quad = I(X_lU_l;Y_l|Q_l) - I(X_l;Y_l|Q_lU_l) - I(X_lU_l;Z_l|Q_l) + I(X_l;Z_l|Q_lU_l) \\
&\quad = I(X_l;Y_l|Q_l) + I(U_l;Y_l|Q_lX_l) - I(X_l;Y_l|U_l) \\
&\qquad - I(X_l;Z_l|Q_l) - I(U_l;Z_l|Q_lX_l) + I(X_l;Z_l|U_l) \\
&\quad \overset{(a)}{=} I(X_l;Y_l|Q_l) - I(X_l;Y_l|U_l) - I(X_l;Z_l|Q_l) + I(X_l;Z_l|U_l) \\
&\quad \leq I(X_l;Y_l|Q_l) - I(X_l;Y_l|U_l) - I(X_l;Z_l|Q_l) + I(X_l;Y_lZ_l|U_l) \\
&\quad = I(X_l;Y_l|Q_l) - I(X_l;Y_l|U_l) - I(X_l;Z_l|Q_l) + I(X_l;Y_l|U_l) + I(X_l;Z_l|U_lY_l) \\
&\quad \overset{(b)}{\leq} I(X_l;Y_l|Q_l) - I(X_l;Z_l|Q_l)
\end{aligned}
\tag{21}
$$

where $(a)$ follows because $I(U_l;Y_l|Q_lX_l) = 0$ and $I(U_l;Z_l|Q_lX_l) = 0$ due to the Markov chain condition $Q_l \to U_l \to X_l \to (Y_lZ_l)$, and $(b)$ follows because $I(X_l;Z_l|U_lY_l) = 0$ due to the degraded condition (15).

By applying the bounds (20) and (21) to the bound on $R_1$ given in (7), we obtain the bound on $R_1$ given in (18). This concludes the proof of the converse. □

# 3 Parallel Gaussian BCCs

In this section, we study the parallel Gaussian BCCs, where the channel outputs at receivers 1 and 2 are corrupted by additive Gaussian noise terms. The channel input-output relationship is given by

$$
Y_{li} = X_{li} + W_{li}, \qquad Z_{li} = X_{li} + V_{li}, \qquad \text{for } \; l = 1, \ldots, L \tag{22}
$$

where $i$ is the time index. For $l = 1, \ldots, L$, the noise processes $\{W_{li}\}$ and $\{V_{li}\}$ are independent identically distributed (i.i.d.) with the components being zero mean Gaussian random variables with the variances $\mu_l^2$ and $\nu_l^2$, respectively. We assume $\mu_l^2 < \nu_l^2$ for $l \in A$ and $\mu_l^2 \geq \nu_l^2$ for $l \in A^c$. The channel input sequence $X_{[1,L]}^n$ is subject to the average power constraints $P$, i.e.,

$$
\frac{1}{n} \sum_{i=1}^{n} \sum_{l=1}^{L} X_{li}^2 \leq P. \tag{23}
$$

We now apply Lemma 1 to obtain the secrecy capacity region of the parallel Gaussian BCC. It can be seen from (22) that the subchannels of the parallel Gaussian BCC are not physically degraded. We consider the following subchannel:

$$\begin{aligned} &Y_{li} = X_{li} + W_{li}, \qquad Z_{li} = X_{li} + W_{li} + V'_{li}, \qquad \text{for } l \in A; \\ &Y_{li} = X_{li} + V_{li} + W'_{li}, \qquad Z_{li} = X_{li} + V_{li}, \qquad \text{for } l \in A^c; \end{aligned} \tag{24}$$

where $\{W'_{li}\}$ and $\{V'_{li}\}$ are i.i.d. random processes with components being zero mean Gaussian random variables with the variances $\mu_l^2 - \nu_l^2$ (for $l \in A^c$) and $\nu_l^2 - \mu_l^2$ (for $l \in A$), respectively. Moreover, $\{V'_{li}\}$ is independent of $\{W_{li}\}$ and $\{W'_{li}\}$ is independent of $\{V_{li}\}$. It can be seen that the channel defined in (24) has physically degraded subchannels. This channel has the same marginal distributions $p(y|x)$ and $p(z|x)$ as the parallel Gaussian BCC defined in (22). Hence by Lemma 1, the two channels have the same secrecy capacity region.

For the channel defined in (24), we can apply Corollary 3 to obtain the secrecy capacity region. In particular, the degradedness of the subchannels allows the use of the entropy power inequality in the proof of the converse. The secrecy capacity region obtained for this channel also applies to the parallel Gaussian BCC defined in (22), and is presented in the following theorem. The details of the proof is relegated to Appendix B.

**Theorem 2.** *The secrecy capacity region of the parallel Gaussian BCC is*

$$\mathcal{C}_s^g = \bigcup_{\underline{p} \in \mathcal{P}} \left\{ \begin{array}{l} (R_0, R_1) : \\ R_0 \le \min \left\{ \displaystyle\sum_{l \in A} \frac{1}{2} \log \left( 1 + \frac{p_{l0}}{\mu_l^2 + p_{l1}} \right) + \sum_{l \in A^c} \frac{1}{2} \log \left( 1 + \frac{p_{l0}}{\mu_l^2} \right), \right. \\ \qquad\qquad \left. \displaystyle\sum_{l \in A} \frac{1}{2} \log \left( 1 + \frac{p_{l0}}{\nu_l^2 + p_{l1}} \right) + \sum_{l \in A^c} \frac{1}{2} \log \left( 1 + \frac{p_{l0}}{\nu_l^2} \right) \right\} \\ R_1 \le \displaystyle\sum_{l \in A} \left[ \frac{1}{2} \log \left( 1 + \frac{p_{l1}}{\mu_l^2} \right) - \frac{1}{2} \log \left( 1 + \frac{p_{l1}}{\nu_l^2} \right) \right] \end{array} \right\}. \tag{25}$$

*where $\underline{p}$ is the power allocation vector, which consists of $(p_{l0}, p_{l1})$ for $l \in A$ and $p_{l0}$ for $l \in A^c$ as components. The set $\mathcal{P}$ includes all power allocation vectors $\underline{p}$ that satisfy the power constraint* (23), *i.e.,*

$$\mathcal{P} := \left\{ \underline{p} : \sum_{l \in A} [p_{l0} + p_{l1}] + \sum_{l \in A^c} p_{l0} \le P \right\}. \tag{26}$$

Note that $\underline{p}$ indicates the power allocation among all subchannels. For $l \in A$, since the source node transmits both common and confidential messages, $p_{l0}$ and $p_{l1}$ indicate the powers allocated to transmit the common and private messages, respectively. For $l \in A^c$, the source transmit only the common message, and $p_{l0}$ indicates the power to transmit the common message.

If $L = 1$, the parallel Gaussian BCC becomes the Gaussian BCC. The following secrecy capacity region of the Gaussian BCC follows directly from Theorem 2.

**Corollary 4.** *The secrecy capacity region of the Gaussian BCC is*

$$\mathcal{C}_s = \bigcup_{0 \le \beta \le 1} \left\{ \begin{array}{l} (R_0, R_1): \\ R_0 \le \min \left\{ \frac{1}{2} \log\left(1 + \frac{(1-\beta)P}{\mu^2 + \beta P}\right), \frac{1}{2} \log\left(1 + \frac{(1-\beta)P}{\nu^2 + \beta P}\right) \right\} \\ R_1 \le \left[ \frac{1}{2} \log\left(1 + \frac{\beta P}{\mu^2}\right) - \frac{1}{2} \log\left(1 + \frac{\beta P}{\nu^2}\right) \right]^+ \end{array} \right\} \tag{27}$$

*where* $(x)^+ = x$ *if* $x > 0$ *and* $(x)^+ = 0$ *if* $x \le 0$.

To characterize the secrecy capacity region of the parallel Gaussian BCC given in (25), we need to characterize every boundary point and the corresponding power allocation vector that achieves this boundary point. It is clear that the secrecy capacity region given in (25) is convex due to the converse proof in Appendix B. Hence the boundary of the secrecy capacity region can be characterized as follows. For every point $(R_0^*, R_1^*)$ on the boundary, there exist $\gamma_0 > 0$ and $\gamma_1 > 0$ such that $(R_0^*, R_1^*)$ is the solution to the following optimization problem

$$\max_{(R_0, R_1) \in \mathcal{C}_s^g} \Big[ \gamma_0 R_0 + \gamma_1 R_1 \Big]. \tag{28}$$

Therefore, the power allocation $\underline{p}^*$ that achieves the boundary point $(R_0^*, R_1^*)$ is the solution to the following optimization problem

$$\begin{aligned} &\max_{\underline{p} \in \mathcal{P}} \Big[ \gamma_0 R_0(\underline{p}) + \gamma_1 R_1(\underline{p}) \Big] \\ &\quad = \max_{\underline{p} \in \mathcal{P}} \Big[ \gamma_0 \min \big\{ R_{01}(\underline{p}), R_{02}(\underline{p}) \big\} + \gamma_1 R_1(\underline{p}) \Big] \end{aligned} \tag{29}$$

where $R_0(\underline{p})$ and $R_1(\underline{p})$ indicate the bounds on $R_0$ and $R_1$ in (25). We further define $R_{01}(\underline{p})$ and $R_{02}(\underline{p})$ to be the two terms over which the minimization in $R_0(\underline{p})$ is taken, i.e., $R_0(\underline{p}) = \min\{R_{01}(\underline{p}), R_{02}(\underline{p})\}$. The optimization (29) serves as a complete characterization of the boundary of the secrecy capacity region of the parallel Gaussian BCC. The solution to (29) provides the power allocations that achieve the boundary of the secrecy capacity region. Our goal now is to solve the optimization problem (29).

The optimization problem (29) is a max-min optimization, and can be solved by the approach used in [47]. The main idea is contained in Proposition 1 in [47], which is stated in the following lemma.

**Lemma 2.** *The optimal $\underline{p}^*$ that solves* (29) *falls into one of the following three cases:*

$$\begin{aligned}
&\textit{Case 1: } \underline{p}^* \textit{ maximizes } \gamma_0 R_{01}(\underline{p}) + \gamma_1 R_1(\underline{p}), \ \textit{and } R_{01}(\underline{p}^*) < R_{02}(\underline{p}^*);\\
&\textit{Case 2: } \underline{p}^* \textit{ maximizes } \gamma_0 R_{02}(\underline{p}) + \gamma_1 R_1(\underline{p}), \ \textit{and } R_{01}(\underline{p}^*) > R_{02}(\underline{p}^*);\\
&\textit{Case 3: } \underline{p}^* \textit{ maximizes } \gamma_0\Big(\alpha R_{01}(\underline{p}) + \bar{\alpha} R_{02}(\underline{p})\Big) + \gamma_1 R_1(\underline{p}),\\
&\qquad\qquad \textit{where } 0 \le \alpha \le 1 \textit{ is such that } R_{01}(\underline{p}^*) = R_{02}(\underline{p}^*), \ \textit{and } \bar{\alpha} = 1-\alpha.
\end{aligned} \tag{30}$$

By applying Lemma 2, we obtain the optimal power allocation $\underline{p}^*$ that solves (29).

**Theorem 3.** *The optimal power allocation vector $\underline{p}^*$ that solves* (29) *and hence achieves the boundary of the secrecy capacity region of the parallel Gaussian BCC has one of the following three forms.*

*Case 1: $\underline{p}^* = \underline{p}^{(1)}$ if the following $\underline{p}^{(1)}$ satisfies $R_{01}\left(\underline{p}^{(1)}\right) < R_{02}\left(\underline{p}^{(1)}\right)$.*

$$\begin{aligned}
&\textit{For } l \in A, \ \textit{if } \frac{\gamma_1}{\gamma_0} > \frac{\nu_l^2}{\nu_l^2 - \mu_l^2},\\
&\qquad p_{l0}^{(1)} = \left(\frac{\gamma_0}{2\lambda \ln 2} - \left(\frac{\gamma_1}{\gamma_0} - 1\right)(\nu_l^2 - \mu_l^2)\right)^+,\\
&\qquad p_{l1}^{(1)} = \left(\min\left\{\frac{1}{2}\sqrt{(\nu_l^2 - \mu_l^2)\left(\nu_l^2 - \mu_l^2 + \frac{2\gamma_1}{\lambda \ln 2}\right)} - \frac{1}{2}\left(\mu_l^2 + \nu_l^2\right),\right.\right.\\
&\qquad\qquad\qquad \left.\left.\frac{\gamma_1}{\gamma_0}(\nu_l^2 - \mu_l^2) - \nu_l^2\right\}\right)^+\\
&\qquad \textit{if } \frac{\gamma_1}{\gamma_0} \le \frac{\nu_l^2}{\nu_l^2 - \mu_l^2}, \quad p_{l0}^{(1)} = \left(\frac{\gamma_0}{2\lambda \ln 2} - \mu_l^2\right)^+, \quad p_{l1}^{(1)} = 0\\
&\textit{For } l \in A^c, \ p_{l0}^{(1)} = \left(\frac{\gamma_0}{2\lambda \ln 2} - \mu_l^2\right)^+
\end{aligned} \tag{31}$$

*where $\lambda$ is chosen to satisfy the power constraint*

$$\sum_{l \in A} [p_{l0} + p_{l1}] + \sum_{l \in A^c} p_{l0} \le P. \tag{32}$$

*Case 2:* $\underline{p}^* = \underline{p}^{(2)}$ *if the following* $\underline{p}^{(2)}$ *satisfies* $R_{01}\left(\underline{p}^{(2)}\right) > R_{02}\left(\underline{p}^{(2)}\right)$.

$$
\begin{aligned}
&\textit{For } \ l \in A, \ \textit{if } \ \frac{\gamma_1}{\gamma_0} > \frac{\mu_l^2}{\nu_l^2 - \mu_l^2}, \\
&\qquad p_{l0}^{(2)} = \left(\frac{\gamma_0}{2\lambda \ln 2} - \left(\frac{\gamma_1}{\gamma_0} + 1\right)\left(\nu_l^2 - \mu_l^2\right)\right)^+, \\
&\qquad p_{l1}^{(2)} = \left(\min\left\{\frac{1}{2}\sqrt{\left(\nu_l^2 - \mu_l^2\right)\left(\nu_l^2 - \mu_l^2 + \frac{2\gamma_1}{\lambda \ln 2}\right)} - \frac{1}{2}\left(\mu_l^2 + \nu_l^2\right),\right.\right. \\
&\qquad\qquad \left.\left.\frac{\gamma_1}{\gamma_0}(\nu_l^2 - \mu_l^2) - \mu_l^2\right\}\right)^+ \\
&\qquad \textit{if } \ \frac{\gamma_1}{\gamma_0} \le \frac{\mu_l^2}{\nu_l^2 - \mu_l^2}, \quad p_{l0}^{(2)} = \left(\frac{\gamma_0}{2\lambda \ln 2} - \nu_l^2\right)^+, \quad p_{l1}^{(1)} = 0 \\
&\textit{For } \ l \in A^c, \ p_{l0}^{(2)} = \left(\frac{\gamma_0}{2\lambda \ln 2} - \nu_l^2\right)^+
\end{aligned}
\tag{33}
$$

*where* $\lambda$ *is chosen to satisfy the power constraint defined in* (32).

*Case 3:* $\underline{p}^* = \underline{p}^{(\alpha)}$ *if there exists* $0 \le \alpha \le 1$ *such that the following* $\underline{p}^{(\alpha)}$ *satisfies* $R_{01}\left(\underline{p}^{(\alpha)}\right) = R_{02}\left(\underline{p}^{(\alpha)}\right)$.

$$
\begin{aligned}
&\textit{For } \ l \in A, \ \textit{if } \ \frac{\gamma_1}{\gamma_0} > \frac{\alpha\nu_l^2 + \bar{\alpha}\mu_l^2}{\nu_l^2 - \mu_l^2}, \\
&\qquad p_{l0}^{(\alpha)} = \left(\frac{1}{2}\sqrt{\left(\nu_l^2 - \mu_l^2 - \frac{\gamma_0}{2\ln 2\lambda}\right)^2 + \frac{2\alpha\gamma_0}{\lambda \ln 2}(\nu_l^2 - \mu_l^2)} + \frac{\gamma_0}{4\ln 2\lambda}\right. \\
&\qquad\qquad \left. - \left(\frac{\gamma_1}{\gamma_0} - \alpha + \frac{1}{2}\right)\left(\nu_l^2 - \mu_l^2\right)\right)^+, \\
&\qquad p_{l1}^{(\alpha)} = \left(\min\left\{\frac{1}{2}\sqrt{\left(\nu_l^2 - \mu_l^2\right)\left(\nu_l^2 - \mu_l^2 + \frac{2\gamma_1}{\lambda \ln 2}\right)} - \frac{1}{2}\left(\mu_l^2 + \nu_l^2\right),\right.\right. \\
&\qquad\qquad \left.\left.\frac{\gamma_1}{\gamma_0}(\nu_l^2 - \mu_l^2) - (\alpha\nu_l^2 + \bar{\alpha}\mu_l^2)\right\}\right)^+ \\
&\qquad \textit{if } \ \frac{\gamma_1}{\gamma_0} \le \frac{\alpha\nu_l^2 + \bar{\alpha}\mu_l^2}{\nu_l^2 - \mu_l^2}, \\
&\qquad p_{l0}^{(\alpha)} = \left(\frac{1}{2}\sqrt{\left(\nu_l^2 - \mu_l^2 - \frac{\gamma_0}{2\ln 2\lambda}\right)^2 + \frac{2\alpha\gamma_0}{\lambda \ln 2}(\nu_l^2 - \mu_l^2)} - \frac{1}{2}\left(\mu_l^2 + \nu_l^2 - \frac{\gamma_0}{2\ln 2\lambda}\right)\right)^+, \\
&\qquad p_{l1}^{(\alpha)} = 0 \\
&\textit{For } \ l \in A^c, \ p_{l0}^{(\alpha)} = \left(\frac{1}{2}\sqrt{\left(\nu_l^2 - \mu_l^2 - \frac{\gamma_0}{2\ln 2\lambda}\right)^2 + \frac{2\alpha\gamma_0}{\lambda \ln 2}(\nu_l^2 - \mu_l^2)} - \frac{1}{2}\left(\mu_l^2 + \nu_l^2 - \frac{\gamma_0}{2\ln 2\lambda}\right)\right)^+
\end{aligned}
\tag{34}
$$

*where $\lambda$ is chosen to satisfy the power constraint defined in* (32).

The proof of Theorem 3 is relegated to Appendix C. Based on Theorem 3, we provide the following algorithm to find the optimal $\underline{p}^*$.

| Algorithm to search $\underline{p}^*$ that solves (29) | |
|---|---|
| Step 1. | Find $\underline{p}^{(1)}$ given in (31).<br>If $R_{01}\left(\underline{p}^{(1)}\right) < R_{02}\left(\underline{p}^{(1)}\right)$, then $\underline{p}^* = \underline{p}^{(1)}$ and finish.<br>Otherwise, go to Step 2. |
| Step 2. | Find $\underline{p}^{(2)}$ given in (33).<br>If $R_{01}\left(\underline{p}^{(2)}\right) > R_{02}\left(\underline{p}^{(2)}\right)$, then $\underline{p}^* = \underline{p}^{(2)}$ and finish.<br>Otherwise, go to Step 3. |
| Step 3. | For a given $\alpha$, find $\underline{p}^{(\alpha)}$ given in (34).<br>Search over $0 \leq \alpha \leq 1$ to find $\alpha$ that satisfies $R_{01}\left(\underline{p}^{(\alpha)}\right) = R_{02}\left(\underline{p}^{(\alpha)}\right)$.<br>Then $\underline{p}^* = \underline{p}^{(\alpha)}$ and finish. |

# 4 Fading BCCs: Ergodic Secrecy Capacity Region

In this section, we study the fading BCC (see Fig. 4), where the channels from the source node to receivers 1 and 2 are corrupted by multiplicative fading gain processes in addition to additive white Gaussian processes. The channel input-output relationship is given by

$$\begin{aligned} Y_i &= h_{1i} X_i + W_i, \\ Z_i &= h_{2i} X_i + V_i, \end{aligned} \tag{35}$$

where $i$ is the time index, $X_i$ is the channel input at the time instant $i$, and $Y_i$ and $Z_i$ are channel outputs at the time instant $i$ at receivers 1 and 2, respectively. The channel gain coefficients $h_{1i}$ and $h_{2i}$ are proper complex random variables. We define $\underline{h}_i := (h_{1i}, h_{2i})$, and assume $\{\underline{h}_i\}$ is a stationary and ergodic vector random process. The noise processes $\{W_i\}$ and $\{V_i\}$ are zero-mean i.i.d. proper complex Gaussian with $W_i$ and $V_i$ having variances $\mu^2$ and $\nu^2$, respectively. The input sequence $\{X_i\}$ is subject to the average power constraint $P$, i.e., $\frac{1}{n}\sum_{i=1}^{n} \mathrm{E}\left[X_i^2\right] \leq P$.

We assume that the channel state information (i.e., the realization of $\underline{h}_i$) is known at both the transmitter and the receivers instantaneously. The channel state information at the source node can be realized by a reliable feedback from the two receivers, who are supposed to receive information from the source node. Depending on the channel state information, the source node can dynamically change its transmission power to achieve better performance. In this section, we assume that there are no delay constraints on the transmitted messages, and the performance criterion we study, i.e., the secrecy capacity region, is averaged over all channel states and is referred to as the *ergodic* secrecy capacity region.

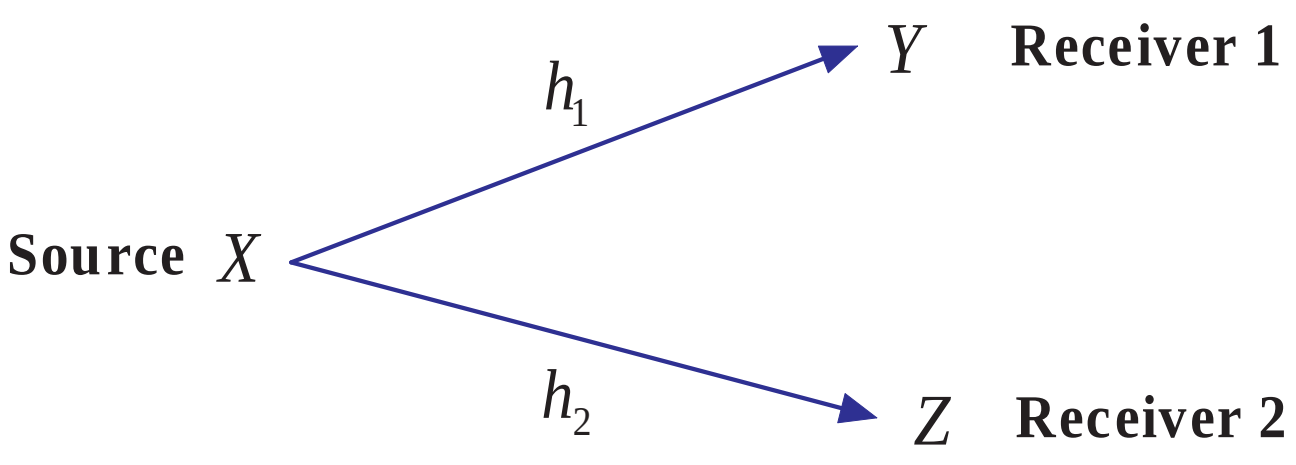

Figure 4: Fading BCC

It can be seen that for a given fading state, i.e., a realization of $\underline{h}_i$, the fading BCC is a Gaussian BCC. Hence the fading BCC can be viewed as a parallel Gaussian BCC with each fading state corresponding to one subchannel. Thus, the following secrecy capacity region of the fading BCC follows from Theorem 2.

**Corollary 5.** *The secrecy capacity region of the fading BCC is*

$$\mathcal{C}_s = \bigcup_{p(\underline{h})\in\mathcal{P}} \left\{ \begin{array}{l} (R_0, R_1): \\ R_0 \le \min\left\{ \mathrm{E}_{\underline{h}\in A} \log\left(1 + \frac{p_0(\underline{h})|h_1|^2}{\mu^2 + p_1(\underline{h})|h_1|^2}\right) + \mathrm{E}_{\underline{h}\in A^c} \log\left(1 + \frac{p_0(\underline{h})|h_1|^2}{\mu^2}\right), \right. \\ \qquad\qquad \left. \mathrm{E}_{\underline{h}\in A} \log\left(1 + \frac{p_0(\underline{h})|h_2|^2}{\nu^2 + p_1(\underline{h})|h_2|^2}\right) + \mathrm{E}_{\underline{h}\in A^c} \log\left(1 + \frac{p_0(\underline{h})|h_2|^2}{\nu^2}\right) \right\} \\ R_1 \le \mathrm{E}_{\underline{h}\in A}\left[\log\left(1 + \frac{p_1(\underline{h})|h_1|^2}{\mu^2}\right) - \log\left(1 + \frac{p_1(\underline{h})|h_2|^2}{\nu^2}\right)\right] \end{array} \right\}. \tag{36}$$

*where* $A := \left\{\underline{h} : \frac{|h_1|^2}{\mu^2} > \frac{|h_2|^2}{\nu^2}\right\}$. *The random vector* $\underline{h} = (h_1, h_2)$ *has the same distribution as the marginal distribution of the process* $\{\underline{h}_i\}$ *at one time instant. The set* $\mathcal{P}$ *is defined as*

$$\mathcal{P} = \left\{p(\underline{h}) : \mathrm{E}_A\left[p_0(\underline{h}) + p_1(\underline{h})\right] + \mathrm{E}_{A^c}[p_0(\underline{h})] \le P\right\}. \tag{37}$$

**Remark 4.** *The secrecy capacity region given in Corollary 5 is established for general fading processes* $\{\underline{h}_i\}$ *where only ergodic and stationary conditions are assumed. The fading process* $\{\underline{h}_i\}$ *can be correlated across time, and is not necessarily Gaussian.*

**Remark 5.** *The secrecy capacity region given in Corollary 5 also applies to the case in which the two component processes* $\{h_{1i}\}$ *and* $\{h_{2i}\}$ *are correlated. However, the secrecy capacity region does depend on the correlation between the two processes. In fact, the average* $E_h$ *in* (36) *needs to be taken over the joint distributions of* $h_{1i}$ *and* $h_{2i}$ *to derive the correct secrecy capacity region.*

This fact can be seen from the following example. We assume only the confidential message is transmitted, i.e., $R_0 = 0$. We also assume that both $h_{1i}$ and $h_{2i}$ take the values of $(0, 1)$ with equal probability $(1/2, 1/2)$. We first consider case 1 in which $h_{1i} = h_{2i}$. From (36) it is clear that $R_1 = 0$, i.e., no secrecy capacity is possible, because the channels to the two receivers are the same for all channel realizations. However, for case 2 in which $h_{1i} = 1 - h_{2i}$, from (36) the secrecy capacity equals $\log\left(1 + \frac{2P}{\mu^2}\right)$, where the power is only allocated to the channel states where $h_{1i} = 1$ and $h_{2i} = 0$. From this example, one can see that the correlation between $h_{1i}$ and $h_{2i}$ affects the secrecy capacity region, although the marginal distributions of $h_{1i}$ and $h_{2i}$ are same for the two cases. We further note that this fact is consistent with Lemma 1 because $h_{1i}$ and $h_{2i}$ are now considered as channel parameters (with each realization corresponding to one subchannel) due to CSI availability at the transmitter. The channel statistics come only from the additive Gaussian noise terms for each subchannel with one realization of $h_{1i}$ and $h_{2i}$.

**Remark 6.** *The secrecy capacity region in Corollary 5 is established for the case with general correlation between the noise variables* $W_i$ *and* $V_i$.

From the bound on $R_1$ in (36), it can be seen that as long as $A$ is not a zero probability event, positive secrecy rate can be achieved. Since fading introduces more randomness to the channel, it is more likely that the channel from the source node to receiver 1 is better than the channel from the source node to receiver 2 for some channel states, and hence positive secrecy capacity can be achieved by exploiting these channel states.

Since the source node is assumed to know the channel state information, it can allocate its power according to the instantaneous channel realization to achieve the best performance, i.e., the boundary of secrecy capacity region. The optimal power allocation that achieves the boundary of the secrecy capacity region for the fading BCC can be derived from Theorem 3 and is given in the following.

**Corollary 6.** *The optimal power allocation $p^*(\underline{h})$ that achieves the boundary of the secrecy capacity region of the fading BCC falls into one of the following three cases.*

*Case 1:* $p^*(\underline{h}) = p^{(1)}(\underline{h})$ *if the following* $p^{(1)}(\underline{h})$ *satisfies* $R_{01}(p^{(1)}(\underline{h})) < R_{02}(p^{(1)}(\underline{h}))$.

$$\begin{aligned}
&\text{For } \underline{h} \in A, \quad \text{if } \frac{\gamma_1}{\gamma_0} > \frac{\nu^2|h_1|^2}{\nu^2|h_1|^2 - \mu^2|h_2|^2}, \\
&\quad p_0^{(1)}(\underline{h}) = \left( \frac{\gamma_0}{\lambda \ln 2} - \left( \frac{\gamma_1}{\gamma_0} - 1 \right) \left( \frac{\nu^2}{|h_2|^2} - \frac{\mu^2}{|h_1|^2} \right) \right)^+, \\
&\quad p_1^{(1)}(\underline{h}) = \left( \min \left\{ \frac{1}{2} \sqrt{ \left( \frac{\nu^2}{|h_2|^2} - \frac{\mu^2}{|h_1|^2} \right) \left( \frac{\nu^2}{|h_2|^2} - \frac{\mu^2}{|h_1|^2} + \frac{4\gamma_1}{\lambda \ln 2} \right) } - \frac{1}{2} \left( \frac{\mu^2}{|h_1|^2} + \frac{\nu^2}{|h_2|^2} \right), \right. \right. \\
&\qquad\qquad \left. \left. \frac{\gamma_1}{\gamma_0} \left( \frac{\nu^2}{|h_2|^2} - \frac{\mu^2}{|h_1|^2} \right) - \frac{\nu^2}{|h_2|^2} \right\} \right)^+ \\
&\qquad\qquad \text{if } \frac{\gamma_1}{\gamma_0} \le \frac{\nu^2|h_1|^2}{\nu^2|h_1|^2 - \mu^2|h_2|^2}, \quad p_0^{(1)}(\underline{h}) = \left( \frac{\gamma_0}{\lambda \ln 2} - \frac{\mu^2}{|h_1|^2} \right)^+, \quad p_1^{(1)}(\underline{h}) = 0 \\
&\text{For } \underline{h} \in A^c, \quad p_0^{(1)}(\underline{h}) = \left( \frac{\gamma_0}{\lambda \ln 2} - \frac{\mu^2}{|h_1|^2} \right)^+.
\end{aligned} \tag{38}$$

*The parameter $\lambda$ is chosen to satisfy the following power constraint*

$$\mathrm{E}_A \left[ p_0(\underline{h}) + p_1(\underline{h}) \right] + \mathrm{E}_{A^c}[p_0(\underline{h})] \le P. \tag{39}$$

*Case 2:* $p^*(\underline{h}) = p^{(2)}(\underline{h})$ *if the following* $p^{(2)}(\underline{h})$ *satisfies* $R_{01}(p^{(2)}(\underline{h})) > R_{02}(p^{(2)}(\underline{h}))$.

$$\begin{aligned}
&\text{For } \underline{h} \in A, \quad \text{if } \frac{\gamma_1}{\gamma_0} > \frac{\mu^2|h_2|^2}{\nu^2|h_1|^2 - \mu^2|h_2|^2}, \\
&\quad p_0^{(2)}(\underline{h}) = \left( \frac{\gamma_0}{\lambda \ln 2} - \left( \frac{\gamma_1}{\gamma_0} + 1 \right) \left( \frac{\nu^2}{|h_2|^2} - \frac{\mu^2}{|h_1|^2} \right) \right)^+, \\
&\quad p_1^{(2)}(\underline{h}) = \left( \min \left\{ \frac{1}{2} \sqrt{ \left( \frac{\nu^2}{|h_2|^2} - \frac{\mu^2}{|h_1|^2} \right) \left( \frac{\nu^2}{|h_2|^2} - \frac{\mu^2}{|h_1|^2} + \frac{4\gamma_1}{\lambda \ln 2} \right) } - \frac{1}{2} \left( \frac{\mu^2}{|h_1|^2} + \frac{\nu^2}{|h_2|^2} \right), \right. \right. \\
&\qquad\qquad \left. \left. \frac{\gamma_1}{\gamma_0} \left( \frac{\nu^2}{|h_2|^2} - \frac{\mu^2}{|h_1|^2} \right) - \frac{\mu^2}{|h_1|^2} \right\} \right)^+ \\
&\qquad\qquad \text{if } \frac{\gamma_1}{\gamma_0} \le \frac{\mu^2|h_2|^2}{\nu^2|h_1|^2 - \mu^2|h_2|^2}, \quad p_0^{(2)}(\underline{h}) = \left( \frac{\gamma_0}{\lambda \ln 2} - \frac{\nu^2}{|h_2|^2} \right)^+, \quad p_1^{(1)}(\underline{h}) = 0 \\
&\text{For } \underline{h} \in A^c, \quad p_0^{(2)}(\underline{h}) = \left( \frac{\gamma_0}{\lambda \ln 2} - \frac{\nu^2}{|h_2|^2} \right)^+
\end{aligned} \tag{40}$$

*where $\lambda$ is chosen to satisfy the power constraint defined in* (39).

*Case 3:* $p^*(\underline{h}) = p^{(\alpha)}(\underline{h})$ *if there exists* $0 \leq \alpha \leq 1$ *such that the following* $p^{(\alpha)}(\underline{h})$ *satisfies* $R_{01}(p^{(\alpha)}(\underline{h})) = R_{02}(p^{(\alpha)}(\underline{h}))$.

$$
\begin{aligned}
&\text{For } \underline{h} \in A, \ \text{ if } \frac{\gamma_1}{\gamma_0} > \frac{\alpha\nu^2|h_1|^2 + \bar{\alpha}\mu^2|h_2|^2}{\nu^2|h_1|^2 - \mu^2|h_2|^2}, \\
&\quad p_0^{(\alpha)}(\underline{h}) = \left( \frac{1}{2}\sqrt{\left(\frac{\nu^2}{|h_2|^2} - \frac{\mu^2}{|h_1|^2} - \frac{\gamma_0}{2\ln\lambda}\right)^2 + \frac{4\alpha\gamma_0}{\lambda\ln 2}\left(\frac{\nu^2}{|h_2|^2} - \frac{\mu^2}{|h_1|^2}\right)} \right. \\
&\qquad\qquad \left. + \frac{\gamma_0}{2\ln 2\lambda} - \left(\frac{\gamma_1}{\gamma_0} - \alpha + \frac{1}{2}\right)\left(\frac{\nu^2}{|h_2|^2} - \frac{\mu^2}{|h_1|^2}\right) \right)^+, \\
&\quad p_1^{(\alpha)}(\underline{h}) = \left( \min\left\{ \frac{1}{2}\sqrt{\left(\frac{\nu^2}{|h_2|^2} - \frac{\mu^2}{|h_1|^2}\right)\left(\frac{\nu^2}{|h_2|^2} - \frac{\mu^2}{|h_1|^2} + \frac{4\gamma_1}{\lambda\ln 2}\right)} - \frac{1}{2}\left(\frac{\mu^2}{|h_1|^2} + \frac{\nu^2}{|h_2|^2}\right), \right. \right. \\
&\qquad\qquad \left. \left. \frac{\gamma_1}{\gamma_0}\left(\frac{\nu^2}{|h_2|^2} - \frac{\mu^2}{|h_1|^2}\right) - \left(\alpha\frac{\nu^2}{|h_2|^2} + \bar{\alpha}\frac{\mu^2}{|h_1|^2}\right) \right\} \right)^+ \\
&\qquad\quad \text{if } \frac{\gamma_1}{\gamma_0} \leq \frac{\alpha\nu^2|h_1|^2 + \bar{\alpha}\mu^2|h_2|^2}{\nu^2|h_1|^2 - \mu^2|h_2|^2}, \\
&\quad p_0^{(\alpha)}(\underline{h}) = \left( \frac{1}{2}\sqrt{\left(\frac{\nu^2}{|h_2|^2} - \frac{\mu^2}{|h_1|^2} - \frac{\gamma_0}{2\ln\lambda}\right)^2 + \frac{4\alpha\gamma_0}{\lambda\ln 2}\left(\frac{\nu^2}{|h_2|^2} - \frac{\mu^2}{|h_1|^2}\right)} \right. \\
&\qquad\qquad \left. - \frac{1}{2}\left(\frac{\mu^2}{|h_1|^2} + \frac{\nu^2}{|h_2|^2} - \frac{\gamma_0}{\lambda\ln 2}\right) \right)^+, \\
&\quad p_1^{(\alpha)}(\underline{h}) = 0 \\
&\text{For } l \in A^c, \ p_0^{(\alpha)}(\underline{h}) = \left( \frac{1}{2}\sqrt{\left(\frac{\nu^2}{|h_2|^2} - \frac{\mu^2}{|h_1|^2} - \frac{\gamma_0}{\lambda\ln 2}\right)^2 + \frac{4\alpha\gamma_0}{\lambda\ln 2}\left(\frac{\nu^2}{|h_2|^2} - \frac{\mu^2}{|h_1|^2}\right)} \right. \\
&\qquad\qquad \left. - \frac{1}{2}\left(\frac{\mu^2}{|h_1|^2} + \frac{\nu^2}{|h_2|^2} - \frac{\gamma_0}{\lambda\ln 2}\right) \right)^+
\end{aligned} \tag{41}
$$

*where $\lambda$ is chosen to satisfy the power constraint defined in* (39).

If the source node does not have common messages for both receivers, and only has confidential messages for receiver 1, the fading BCC becomes the fading wire-tap channel. For this channel, the secrecy capacity is readily obtained from Corollaries 5 and 6.

**Corollary 7.** *The secrecy capacity of the fading wire-tap channel is*

$$C_s = \max_{\mathrm{E}_A[p(\underline{h})]\leq P} \mathrm{E}_A\left[\log\left(1+\frac{p(\underline{h})|h_1|^2}{\mu^2}\right) - \log\left(1+\frac{p(\underline{h})|h_2|^2}{\nu^2}\right)\right]. \tag{42}$$

*The optimal power allocation that achieves the secrecy capacity in* (42) *is given by*

$$p^*(\underline{h}) = \begin{cases} \left(\frac{1}{\lambda\ln 2} - \frac{\mu^2}{|h_1|^2}\right)^+, & \text{if } |h_2|^2 = 0; \\ \left(\frac{1}{2}\sqrt{\left(\frac{\nu^2}{|h_2|^2}-\frac{\mu^2}{|h_1|^2}\right)\left(\frac{4}{\lambda\ln 2}-\frac{\mu^2}{|h_1|^2}+\frac{\nu^2}{|h_2|^2}\right)} - \frac{1}{2}\left(\frac{\mu^2}{|h_1|^2}+\frac{\nu^2}{|h_2|^2}\right)\right)^+, & \\ & \text{if } |h_2|^2 > 0,\ \underline{h}\in A; \\ 0, & \text{otherwise} \end{cases} \tag{43}$$

*where* $\lambda$ *is chosen to satisfy the power constraint* $\mathrm{E}_A[p(\underline{h})] = P$.

# 5 Fading BCCs: Outage Performance

In Section 4, we considered the ergodic secrecy capacity region for the fading BCC. In this case, messages can be coded over long block lengths and hence over all channel realizations. This applies to wireless systems in which the transmission delay can be tolerant. In this section, we consider wireless systems in which there is a stringent delay constraint, and messages must be transmitted within a certain time.

We adopt the channel model described in (35). However, we now take the block fading assumption, where the fading coefficients $h_{1i}$ and $h_{2i}$ remain constant over one block and change to another realization in the next block in an ergodic and stationary matter. Moreover, we assume that the block length is large enough such that coding over one block can achieve small probability of error. We assume that the delay constraint is within the block length. Coding over multiple blocks and hence over multiple channel state realizations is not allowed. We also assume that both the transmitter and the receivers know the channel state information.

We use $(\check{R}_0, \check{R}_1)$ to indicate a target rate pair, i.e., the common and confidential messages need to be transmitted to the two receivers at the rates $\check{R}_0$ and $\check{R}_1$, respectively, in each block (each fading state realization). If the target rate pair is not achieved for one block, an outage is claimed. We define the outage probability to be

$$P_{out} = Pr\{(\check{R}_0, \check{R}_1) \notin \mathcal{C}_s(\underline{h}, p(\underline{h}))\} \tag{44}$$

where $\mathcal{C}_s(\underline{h}, p(\underline{h}))$ is the secrecy capacity region for the channel with fading state realization $\underline{h}$, and $p(\underline{h})$ indicates the transmission power used by the source node for this fading state. The source node is able to adapt its transmission power to the instantaneous channel state realization, i.e., $p(\underline{h})$ is a function of $\underline{h}$, because CSI is assumed to be known at the transmitter. We assume that the power constraint applies over a large number of blocks and hence over all fading state realizations (i.e., it is a long term power constraint as in [16]); that is, we assume

$$\mathrm{E}[p(\underline{h})] \leq P. \tag{45}$$

We define the set $\mathcal{P} := \{p(\underline{h}) : \mathrm{E}[p(\underline{h})] \leq P\}$.

It is clear from (44) that the outage probability depends on the power allocation function $p(\underline{h})$. Our goal is to study the power allocation $p^*(\underline{h})$ that minimizes the outage probability, i.e.,

$$\min_{p(\underline{h}) \in \mathcal{P}} P_{out}. \tag{46}$$

To understand this problem, we note that for each channel state the source node knows how much power it needs to use to support the target rate pair. If the power needed is too large, the source node may decide not to transmit and claim outage in order to save power for other channel states that need lower power to support the target rate pair. Hence the source node first allocates power to those channel states that need lower power to support the target rate, and then to those channel states that need higher power to support the target rate until all power is utilized. This suggests that the power allocation is a threshold decision.

To formally solve the optimization problem (46), we apply the approach in [16], where the minimum outage probability of the fading channel without a secrecy constraint was studied. This approach was also applied in [10] to study the minimum outage probability of the fading broadcast channel without a secrecy constraint. In the following, we describe the power allocation that minimizes the outage probability of the fading BCC, i.e., the solution to (46).

For a given block with the fading state realization $\underline{h} = (h_1, h_2)$, the channel we consider is a Gaussian BCC. From Corollary 4, the secrecy capacity region is given as follows.

If $\underline{h} \in A$,

$$\mathcal{C}_s(\underline{h}, p(\underline{h})) = \bigcup_{0\leq\beta\leq 1} \left\{ \begin{array}{l} (R_0, R_1): \\ R_0 \leq \log\left(1 + \frac{(1-\beta)p(\underline{h})|h_2|^2}{\nu^2 + \beta p(\underline{h})|h_2|^2}\right), \\ \\ R_1 \leq \log\left(1 + \frac{\beta p(\underline{h})|h_1|^2}{\mu^2}\right) - \log\left(1 + \frac{\beta p(\underline{h})|h_2|^2}{\nu^2}\right) \end{array} \right\}$$

If $\underline{h} \in A^c$,

$$\mathcal{C}_s(\underline{h}, p(\underline{h})) = \left\{ \begin{array}{l} (R_0, R_1): \\ R_0 \leq \log\left(1 + \frac{p(\underline{h})|h_1|^2}{\mu^2}\right), \\ R_1 = 0 \end{array} \right\} \tag{47}$$

We now use (47) to compute the minimum power that is needed to achieve the target rate pair $(\check{R}_0, \check{R}_1)$. It is clear from (47) that, only when $\underline{h} \in A$ can we possibly achieve a positive $\check{R}_1$. The minimum power needed to achieve $\check{R}_1$ is

$$\beta p(\underline{h}) = \begin{cases} \dfrac{2^{\check{R}_1} - 1}{\frac{|h_1|^2}{\mu^2} - 2^{\check{R}_1}\frac{|h_2|^2}{\nu^2}}, & \text{if } \check{R}_1 < \log\frac{|h_1|^2\nu^2}{|h_2|^2\mu^2} \\ \infty & \text{otherwise} \end{cases} \tag{48}$$

The minimum power to support $\check{R}_0$ is then given by

$$(1-\beta)p(\underline{h}) = \begin{cases} \dfrac{\left(2^{\check{R}_0} - 1\right)\left(\frac{|h_1|^2\nu^2}{|h_2|^2\mu^2} - 1\right)}{\frac{|h_1|^2}{\mu^2} - 2^{\check{R}_1}\frac{|h_2|^2}{\nu^2}}, & \text{if } \check{R}_1 < \log\frac{|h_1|^2\nu^2}{|h_2|^2\mu^2} \\ \infty & \text{otherwise} \end{cases} \tag{49}$$

Hence the minimum power needed to support $(\check{R}_0, \check{R}_1)$ is

$$p^{min}(\underline{h}) = \begin{cases} \dfrac{2^{\check{R}_0}\left(\frac{|h_1|^2\nu^2}{|h_2|^2\mu^2} - 1\right) + 2^{\check{R}_1} - \frac{|h_1|^2\nu^2}{|h_2|^2\mu^2}}{\frac{|h_1|^2}{\mu^2} - 2^{\check{R}_1}\frac{|h_2|^2}{\nu^2}}, & \text{if } \check{R}_1 < \log\frac{|h_1|^2\nu^2}{|h_2|^2\mu^2} \\ \infty & \text{otherwise} \end{cases} \tag{50}$$

For $s > 0$, we define

$$\mathcal{R}(s) = \left\{\underline{h} : p^{min}(\underline{h}) < s\right\}$$
$$\bar{\mathcal{R}}(s) = \left\{\underline{h} : p^{min}(\underline{h}) \leq s\right\} \tag{51}$$

where $p^{min}(\underline{h})$ is given in (50).

The average powers that are needed to support the rate pair $(\check{R}_0, \check{R}_1)$ for the channel states in $\mathcal{R}(s)$ and $\bar{\mathcal{R}}(s)$ are

$$p(s) = \mathrm{E}_{\underline{h}\in\mathcal{R}(s)}\left[p^{min}(\underline{h})\right]$$
$$\bar{p}(s) = \mathrm{E}_{\underline{h}\in\bar{\mathcal{R}}(s)}\left[p^{min}(\underline{h})\right] \tag{52}$$

For the given power constraint $P$, define

$$s^* = \sup\{s : p(s) < P\}$$
$$w^* = \frac{P - p(s^*)}{\bar{p}(s^*) - p(s^*)} \tag{53}$$

By using (50) and applying Lemma 3 and Proposition 4 in [16], we obtain the following optimal power allocation $p^*(\underline{h})$.

**Proposition 1.** *The power allocation $p^*(\underline{h})$ that solves (46), and hence minimizes the outage probability for a given target rate pair $(\check{R}_0, \check{R}_1)$, is given by*

$$p^*(\underline{h}) = \begin{cases} p^{min}(\underline{h}), & \text{if } \underline{h} \in \mathcal{R}(s^*) \\ p^{min}(\underline{h}), & \text{with probability } w^* \text{ if } \underline{h} \in \bar{\mathcal{R}}(s^*)\backslash\mathcal{R}(s^*) \\ 0 & \text{if } \underline{h} \notin \bar{\mathcal{R}}(s^*) \end{cases} \tag{54}$$

*where $p^{min}(\underline{h})$ is given in (50).*

It can be seen that the optimal power allocation $p^*(\underline{h})$ is a threshold solution. The power is first allocated to the fading states that need smaller amounts of power to achieve the target rate pair, and is then allocated to the fading states that need larger amounts of power to achieve the target rate pair. When the total power $P$ is used up by these fading states, no further power is allocated to other states.

In the above problem setting, an outage is claimed if either $\check{R}_0$ or $\check{R}_1$ is not achieved. This results in an optimal solution $p^*(\underline{h})$ that allocates power only to those states for which both $\check{R}_0$ and $\check{R}_1$ can be achieved by relatively small power consumption. However, some channel states may support one target rate with small power consumption, but need a large amount of power to support both rates. These states are unlikely to be allocated power. For example, even if the channels from the source node to both receivers are good to transmit common messages, it may happen that no power is allocated to this fading state if the channel from the source node to receiver 1 is worse than the channel from the source node to receiver 2 so that $\check{R}_1$ cannot be achieved. Sometimes this is not reasonable, because the two messages are independent, and one message should be transmitted whenever the channel is good to transmit it. It should not depend on whether the other message is transmitted or not. Nevertheless, the solution to the problem (46) we have considered is useful if we consider the following two more reasonable problems.

It is clear that the solution to the problem (46) immediately implies the optimal power allocation for the case where only the confidential message is transmitted and only the target rate $\check{R}_1$ is assumed. Now the minimum power to achieve $\check{R}_1$ is given by

$$P^{min}(\underline{h}) = \begin{cases} \dfrac{2^{\check{R}_1} - 1}{\frac{|h_1|^2}{\mu^2} - 2^{\check{R}_1}\frac{|h_2|^2}{\nu^2}}, & \text{if } \check{R}_1 < \log \frac{|h_1|^2 \nu^2}{|h_2|^2 \mu^2} \\ \infty & \text{otherwise} \end{cases} \tag{55}$$

The power allocation that minimizes the outage probability follows from Proposition 1 by using (55) to replace (50) in (54).

We next consider the scenario in which the source node has both common and confidential messages to transmit. We assume that the common message is required to be transmitted at a constant rate $\check{R}_0$ for all channel states, i.e., no outage is allowed for the common rate. This scenario applies to wireless systems where a constant common rate must be satisfied. Since $\check{R}_0$ must be achieved for all channel states, the total power must be large enough to support this rate, i.e.,

$$P \geq P_0 := \mathrm{E}[p_0(\underline{h})] \tag{56}$$

where $p_0(\underline{h})$ is the power that is needed to support the rate $\check{R}_0$ for the channel state $\underline{h}$ and is given by

$$p_0(\underline{h}) = \frac{2^{\check{R}_0} - 1}{\min\left\{\frac{|h_1|^2}{\mu^2}, \frac{|h_2|^2}{\nu^2}\right\}} \tag{57}$$

In addition to the common message, the source node wishes to transmit confidential information to receiver 1 at a target rate $\check{R}_1$ and with as small an outage probability as possible. We note that a similar problem was studied in [10] for the broadcast channel with separate messages for two receivers and the rate to one receiver must be constant for all channel states.

We need to find the power allocation that minimizes the outage probability that the target rate $\check{R}_1$ is not achieved. The optimization problem is summarized as follows:

$$\begin{aligned} \text{Minimize} \quad & Pr\{(\check{R}_0, \check{R}_1\} \notin \mathcal{C}(\underline{h}, p(\underline{h}))\} \\ \text{Subject to} \quad & \mathrm{E}[p(\underline{h})] \leq P \\ & \check{R}_0 \text{ is achieved for all } \underline{h}, \text{ i.e., } p(\underline{h}) \geq p_0(\underline{h}) \end{aligned} \tag{58}$$

We can change the problem (58) to the following equivalent problem

$$\begin{aligned} \text{Minimize} \quad & Pr\{(\check{R}_0, \check{R}_1\} \notin \mathcal{C}(\underline{h}, p(\underline{h}))\} \\ \text{Subject to} \quad & \mathrm{E}[\Delta p(\underline{h})] \leq P - P_0 \end{aligned} \tag{59}$$

where $\Delta p(\underline{h}) = p(\underline{h}) - p_0(\underline{h})$ is the difference between the power needed to support $(\check{R}_0, \check{R}_1)$ and the power needed to support $\check{R}_0$ only.

It can be seen that the problem (59) is the same as the problem (46) with $p(\underline{h})$ in (46) being replaced by $\Delta p(\underline{h})$. Thus the optimal $\Delta p^*(\underline{h})$ can be derived from Proposition 1 with $p^{min}(\underline{h})$ in (54) being replaced by $\Delta p^{min}(\underline{h})$, which is the minimum difference between the power needed to support $(\check{R}_0, \check{R}_1)$ and the power needed to support $\check{R}_0$ and is given by

$$\Delta p^{min}(\underline{h}) = \begin{cases} \dfrac{2^{\check{R}_0}\left(2^{\check{R}_1} - 1\right)}{\frac{|h_1|^2}{\mu^2} - 2^{\check{R}_1}\frac{|h_2|^2}{\nu^2}}, & \text{if } \check{R}_1 < \log \frac{|h_1|^2\nu^2}{|h_2|^2\mu^2} \\ \infty & \text{otherwise.} \end{cases} \tag{60}$$

# 6 Numerical Results

In this section, we provide numerical results to demonstrate the ergodic and outage performance for the fading BCC.

We first consider the Rayleigh fading BCC, where $h_1$ and $h_2$ are zero mean proper complex Gaussian random variables. Hence $|h_1|^2$ and $|h_2|^2$ are exponentially distributed with parameters $\sigma_1$ and $\sigma_2$. We assume the source power $P = 5$ dB, and fix $\sigma_1 = 1$. In Fig. 5, we plot the boundaries of the secrecy capacity regions corresponding to $\sigma_2 = 0.4, 0.7, 1$, respectively. It can be seen that as $\sigma_2$ decreases, the secrecy rate $R_1$ of the confidential message improves, but the rate $R_0$ of the common message decreases. This fact follows because smaller $\sigma_2$ implies worse channel from the source node to receiver 2. Thus, confidential information can be forwarded to receiver 1 at a larger rate. However, the rate of the common information is limited by the worse channel from the source node to receiver 2.

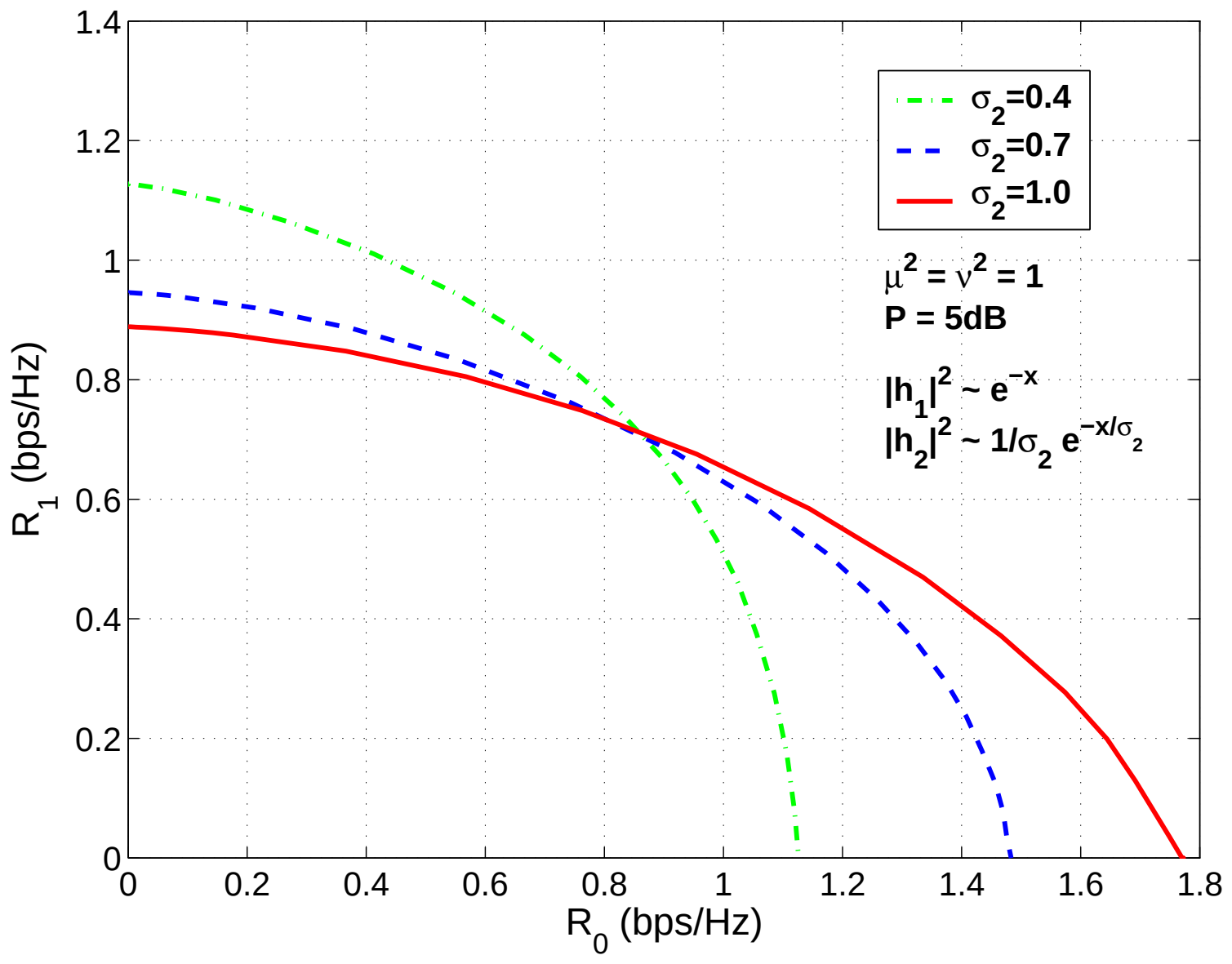

Figure 5: Secrecy capacity regions for a Rayleigh fading BCC

For the Rayleigh fading BCC with $\sigma_1 = 1$ and $\sigma_2 = 0.4$, we plot the boundary of the secrecy capacity region in Fig. 6. The three cases (see Corollary 6) to derive the boundary achieving power allocations are also indicated with the corresponding boundary points. It can be seen that the boundary points with large $R_1$ are achieved by the power allocations derived from Case 1, and are indicated by the line with circle on the graph. The boundary points with large $R_0$ are achieved by the optimal power allocations derived from Case 2, and are indicated by the line with square. Between the boundary points achieved by Case 1 and Case 2, the boundary points are achieved by the power allocations derived from Case 3, and are indicated by the plain solid line.

The intuitive reason how the three cases associate with the boundary points is given as follows. To achieve large secrecy rate $R_1$, most channel states in the set $A$ where receiver 1 has a stronger channel than receiver 2 are used to transmit the confidential message. The common message is hence transmitted mostly over the channel states in the set $A^c$, over which the common rate is limited by the channel from the source node to receiver 1. Thus, power allocation needs to optimize the rate of this channel, and hence the optimal power allocation follows from Case 1. To achieve large $R_0$, the common message is forwarded over the channel states both in $A$ and $A^c$. It can be seen that in average the source node has a much worse channel to receiver 2 than to receiver 1, and hence the channel from the source node to receiver 2 limits the common rate. Power allocation now needs to optimize the rate to receiver 2, and the optimal power allocation follows from Case 2. Between these two cases, power allocation needs to balance the rates to receivers 1 and 2 and hence follows from Case 3.

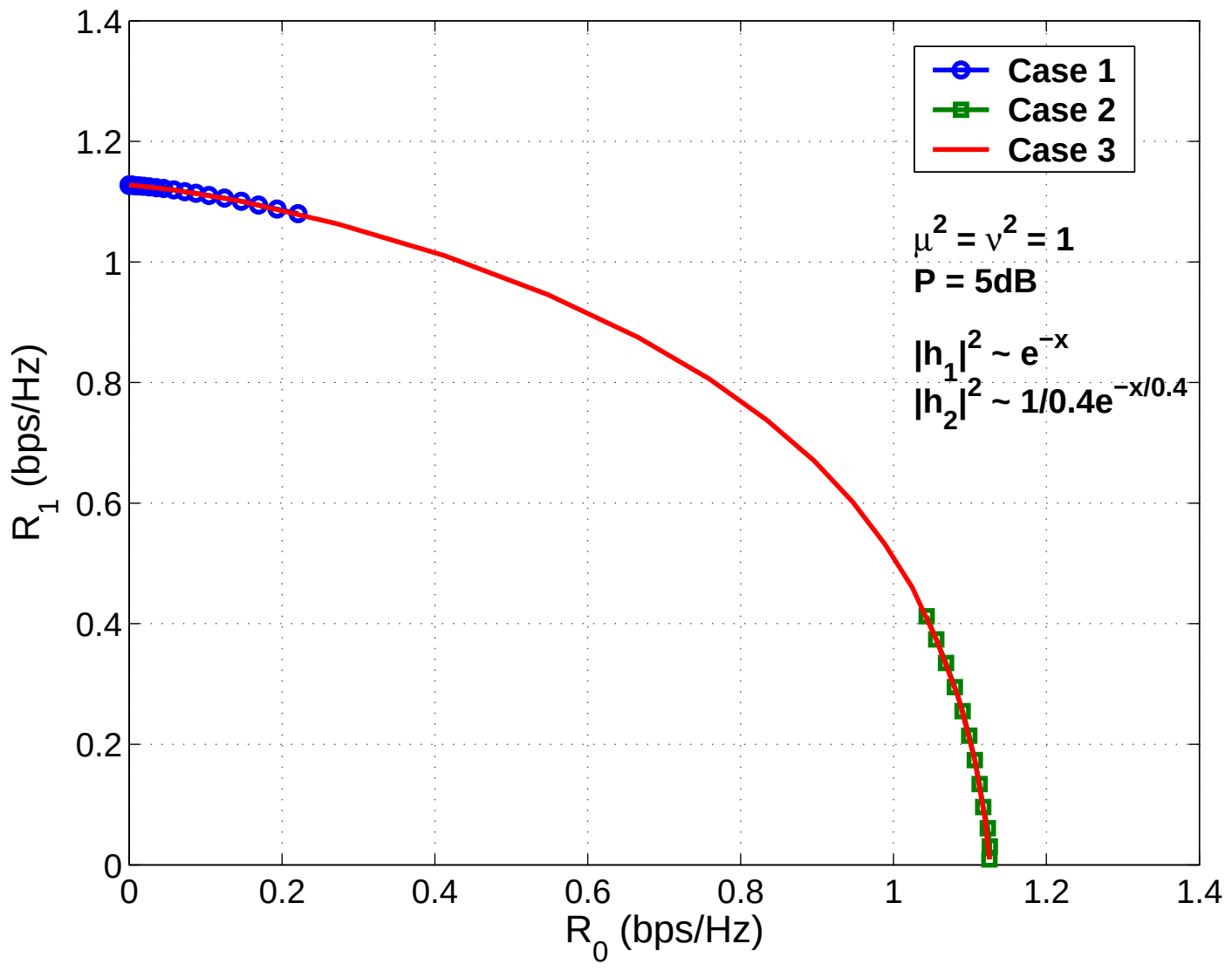

Figure 6: Three cases in power allocation optimization to achieve the boundary of the secrecy capacity region for a Rayleigh fading BCC

We next consider the case in which $R_0 = 0$, i.e., only the confidential message is transmitted from the source node to receiver 1. We assume $\sigma_1 = \sigma_2 = 1$. In Fig. 7 (a), we plot the optimal power allocation $p_1^*(\underline{h})$ as a function of $\underline{h}$. It can be seen from the graph that most of the source power is allocated to the channel states with small $|h_2|^2$. This behavior is shown more clearly in Fig. 7 (b), which plots $p_1^*(\underline{h})$ as a function of $|h_1|^2$ for different values of $|h_2|^2$, and in Fig. 7 (c), which plots $p_1^*(\underline{h})$ as a function of $|h_2|^2$ for different values of $|h_1|^2$. The source node allocates more power to the channel states with larger $|h_1|^2$ to forward more confidential information to the destination node, and allocates less power for the channel states with larger $|h_2|^2$ to prevent receiver 2 to obtain the confidential information. It can also be seen from Fig. 7 (b) and Fig. 7 (c) that the source node transmits only when the channel from the source node to receiver 1 is better than the channel from the source node to receiver 2.

For the case in which $R_0 = 0$, Fig. 8 plots the secrecy capacity achieved by the optimal power allocation, and compares it with the secrecy rate achieved by a uniform power allocation, i.e., allocating the same power for all channel states $\underline{h} \in A$. It can be seen that the uniform power allocation does not provide performance close to the secrecy capacity for the SNRs of interest. This is in contrast to the Rayleigh fading channel without the secrecy constraint, where the uniform power allocation can be close to optimal even for moderate SNRs. This also demonstrates that the exact channel state information is important to achieve higher secrecy rate.

We now consider the outage performance of the Rayleigh fading BCC. We assume $\sigma_1 = 10$ and $\sigma_2 = 0.5$. We first assume the target rate $\check{R}_0 = 0$, i.e., only the confidential message is transmitted. In Fig. 9, we plot the outage probabilities corresponding to different values of the target rate $\check{R}_1$. The outage probability decreases as the target rate $\check{R}_1$ decreases. For a fixed $\check{R}_1$, the outage probability is bounded below by a certain threshold. This fact follows because the outage probability cannot be prevented for those channel states where the channel from the source node to receiver 1 is worse than the channel from the source node to receiver 2. In Fig. 10, we compare the outage probability when $R_0 = 0$ with the outage probability when $R_0 = 0.1$ bps/Hz. It can be seen that even a small positive common rate $\check{R}_0$ can cause large increase in outage probability. In Fig. 11, we compare the outage probability minimized by the power allocation and that achieved by the equal power allocation for all channel states. It can be seen that optimizing power allocation significantly reduces the outage probability.

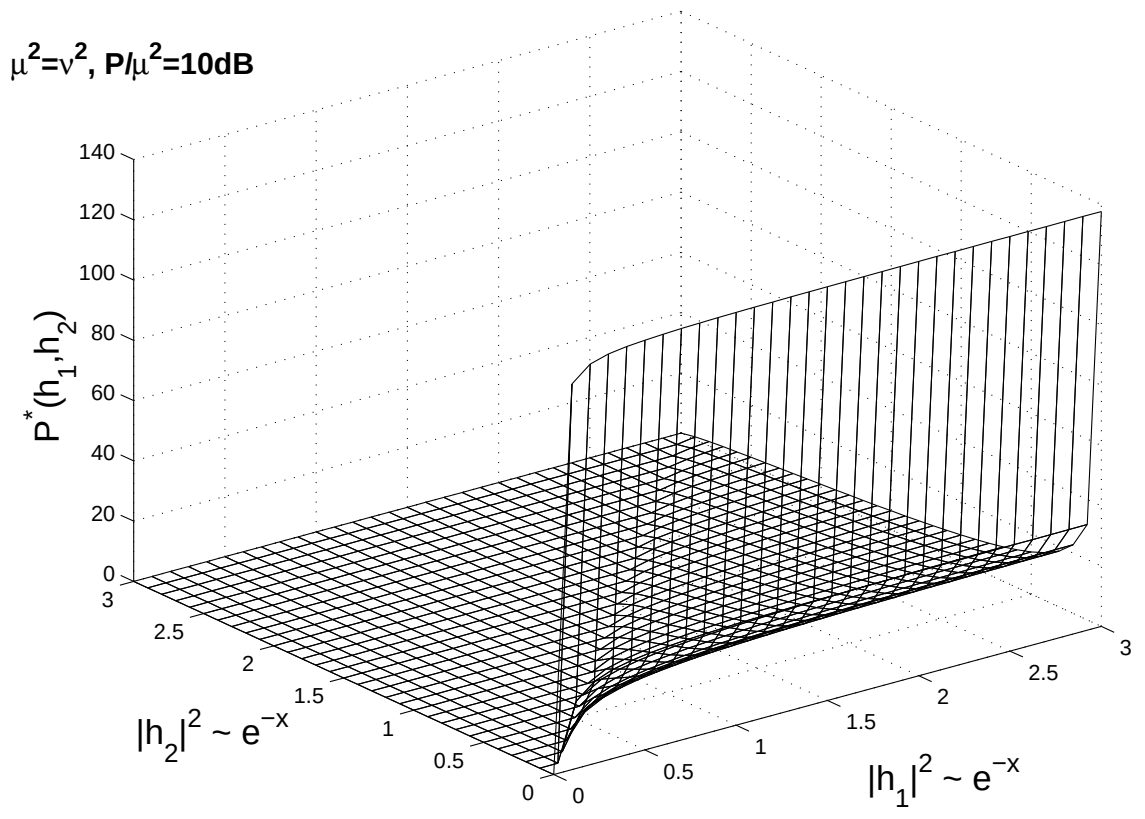

(a): $p_1^*(\underline{h})$ as a function of $(h_1, h_2)$

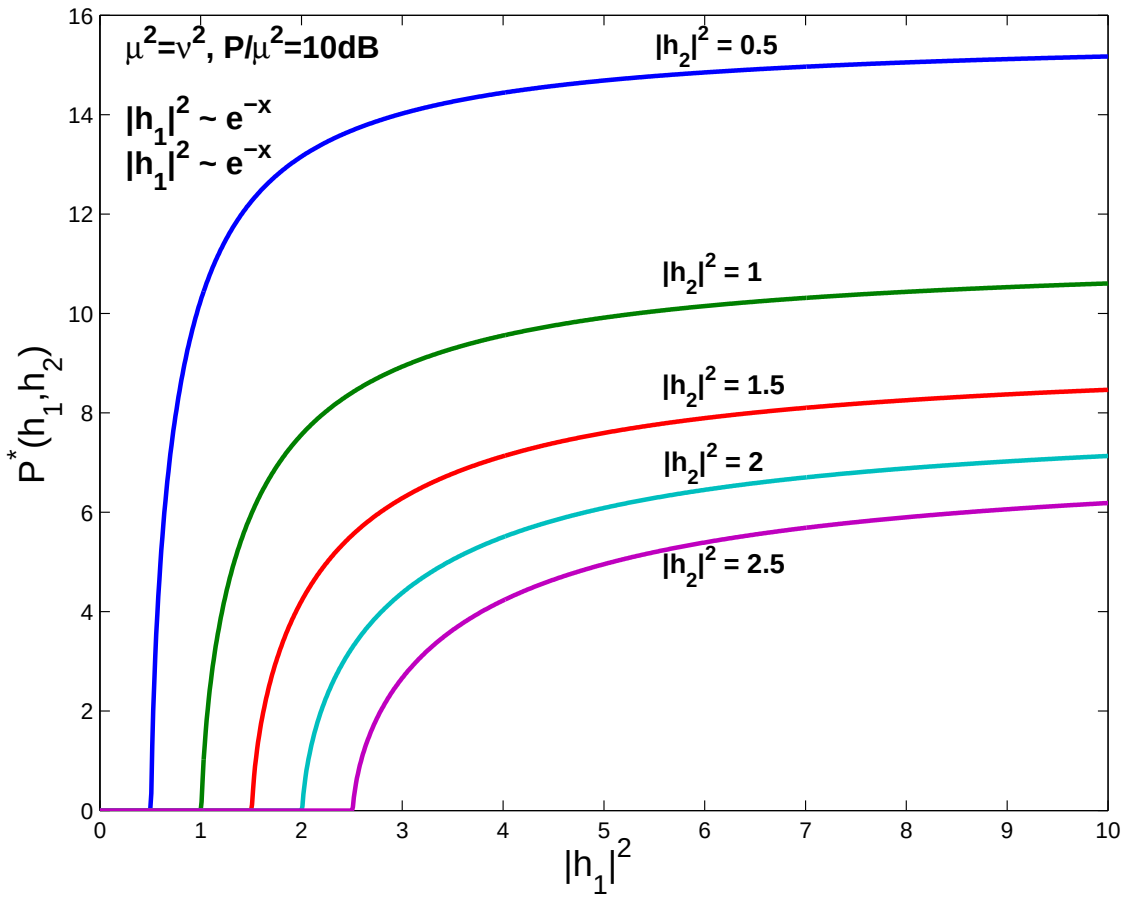

(b): $p_1^*(\underline{h})$ as a function of $|h_1|^2$

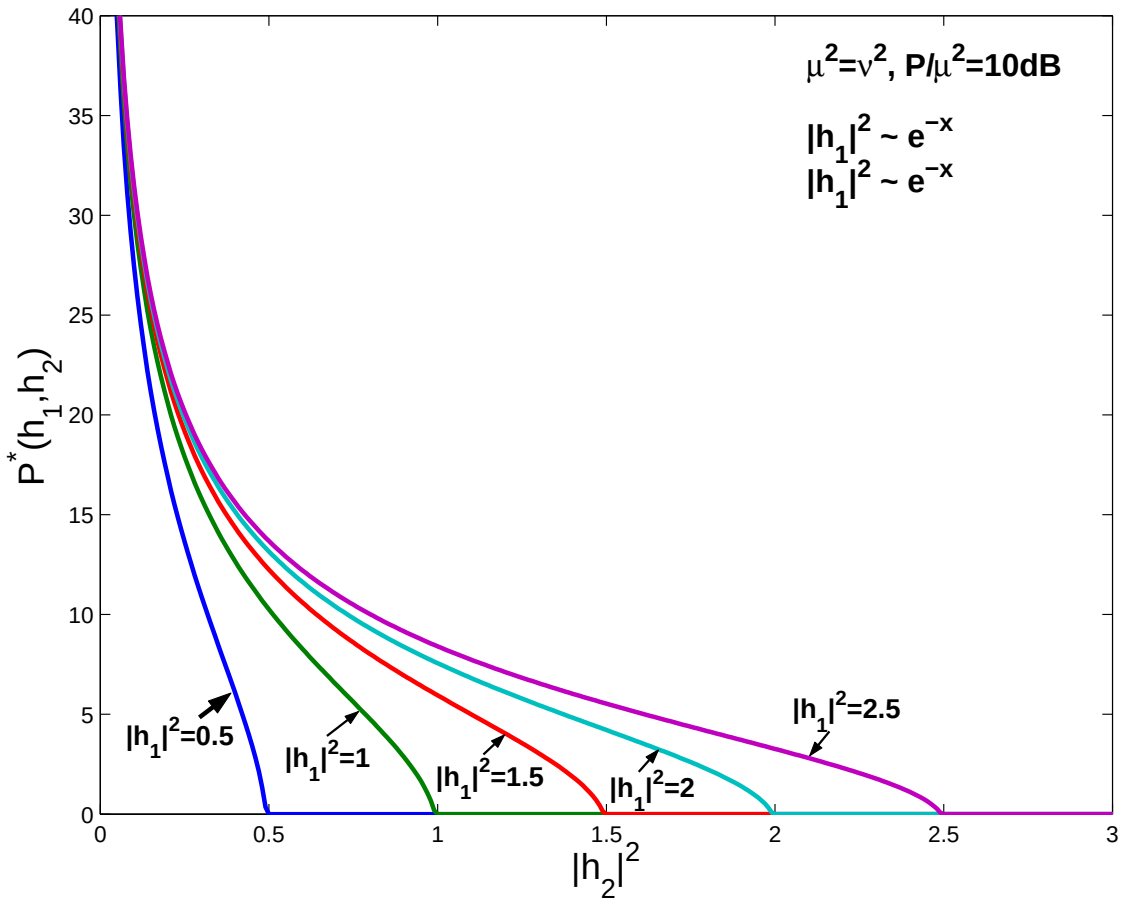

(c): $p_1^*(\underline{h})$ as a function of $|h_2|^2$

Figure 7: Optimal power allocation $p_1^*(\underline{h})$ for a Rayleigh fading BCC with $R_0 = 0$

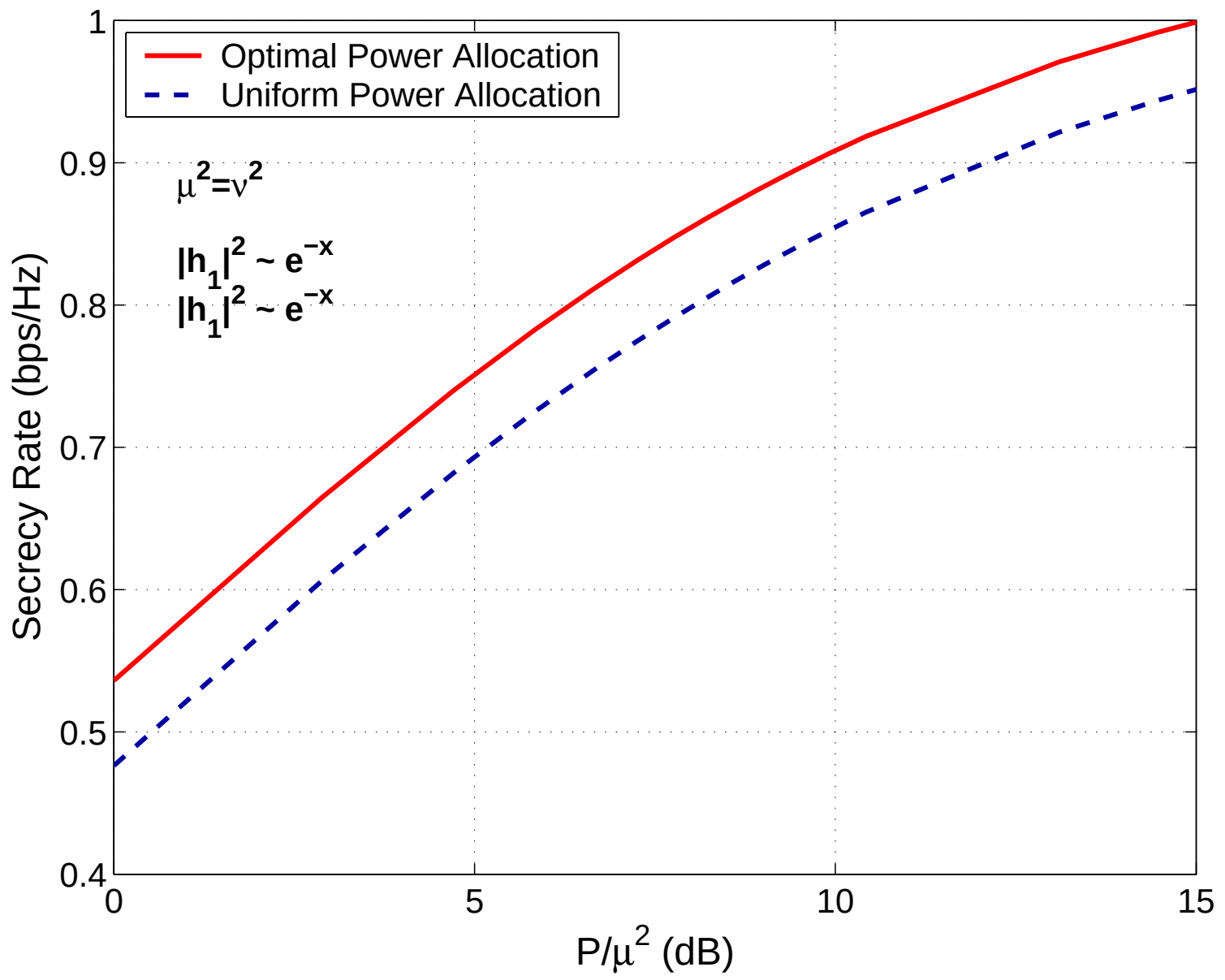

Figure 8: Comparison of secrecy capacity by optimal power allocation with secrecy rate by uniform power allocation for a Rayleigh fading BCC with $R_0 = 0$

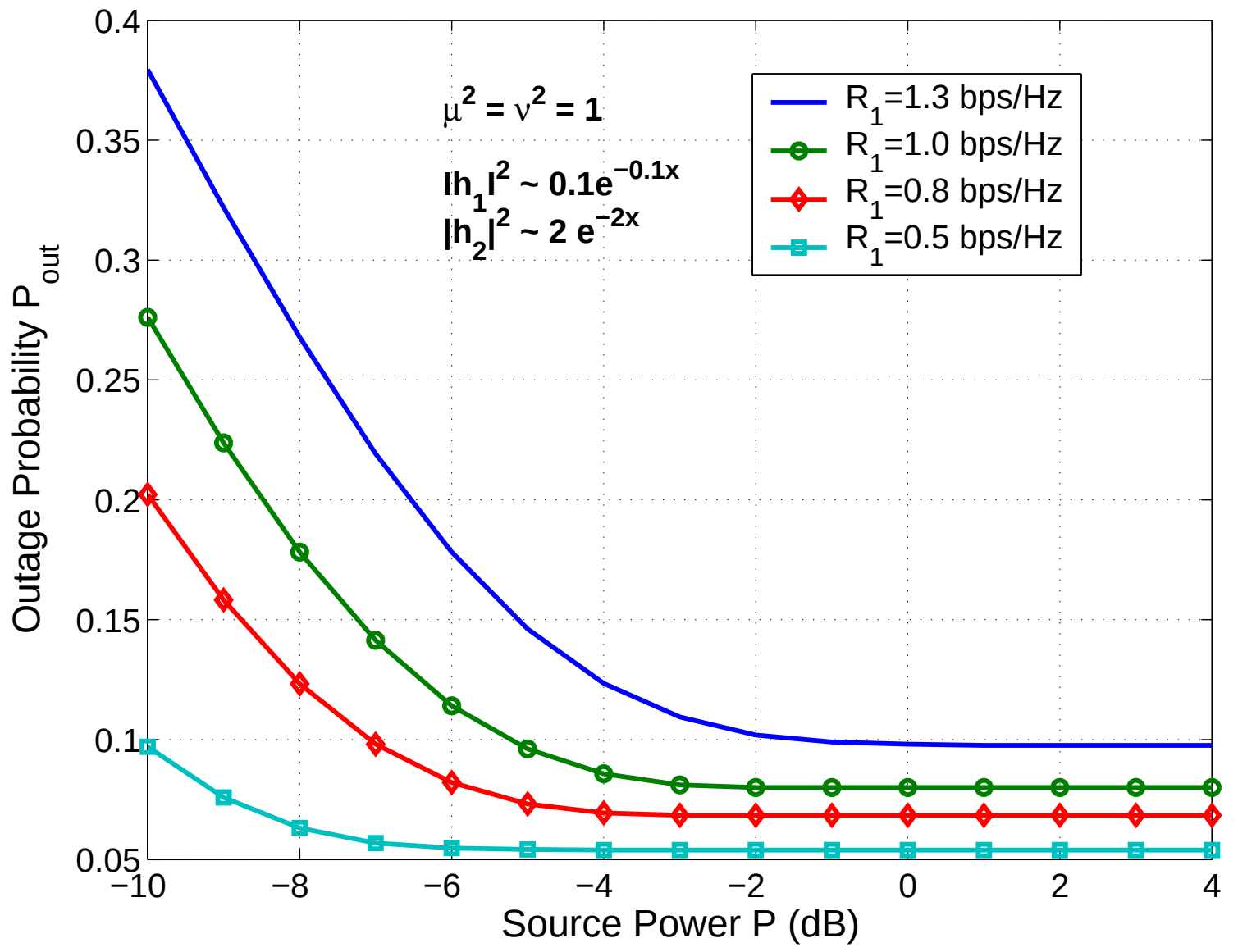

Figure 9: Outage probabilities for a Rayleigh fading BCC with $\check{R}_0 = 0$

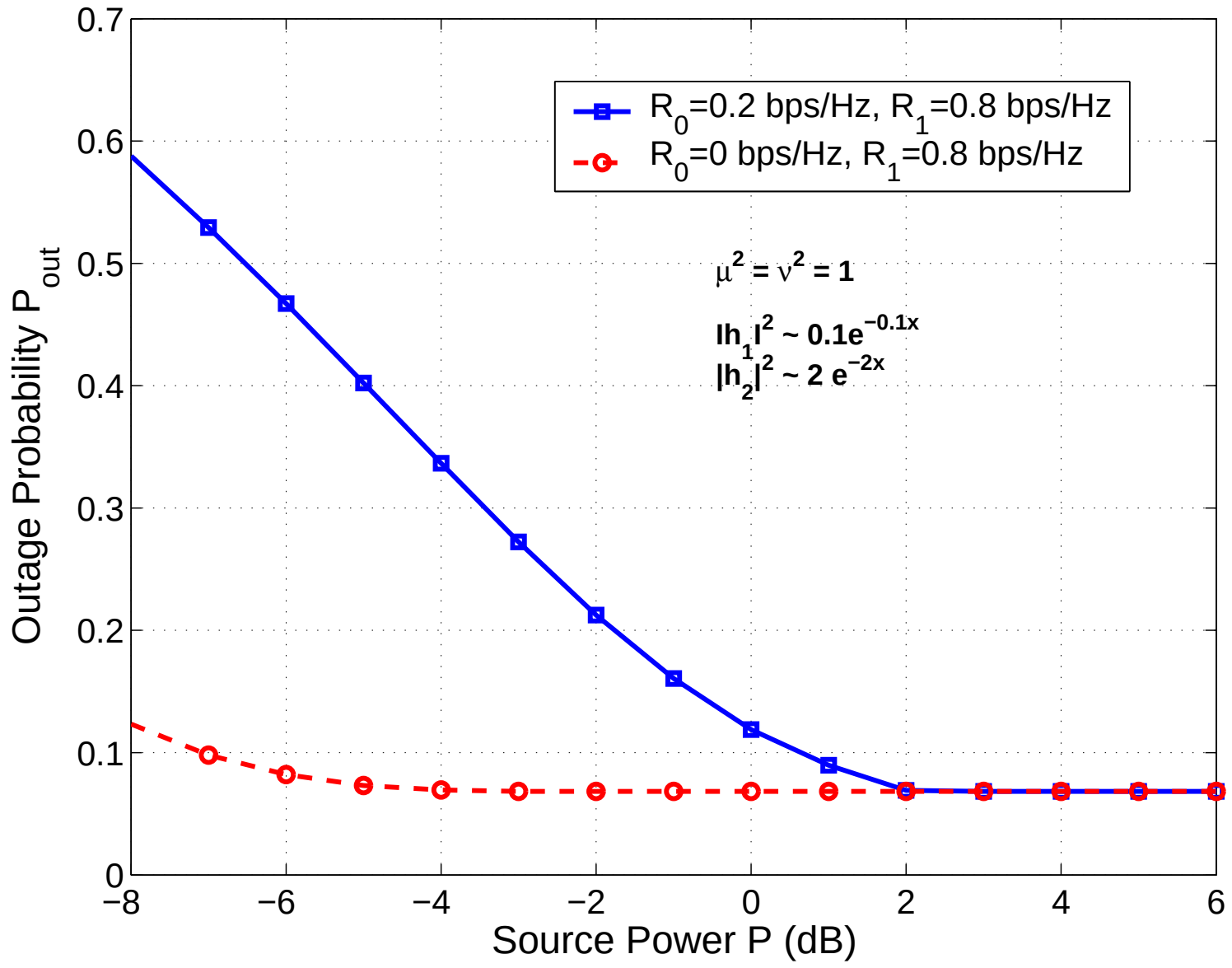

Figure 10: Comparison of outage probabilities when $R_0 = 0$ with those when $R_0 = 0.1$ bps/Hz for a Rayleigh fading BCC

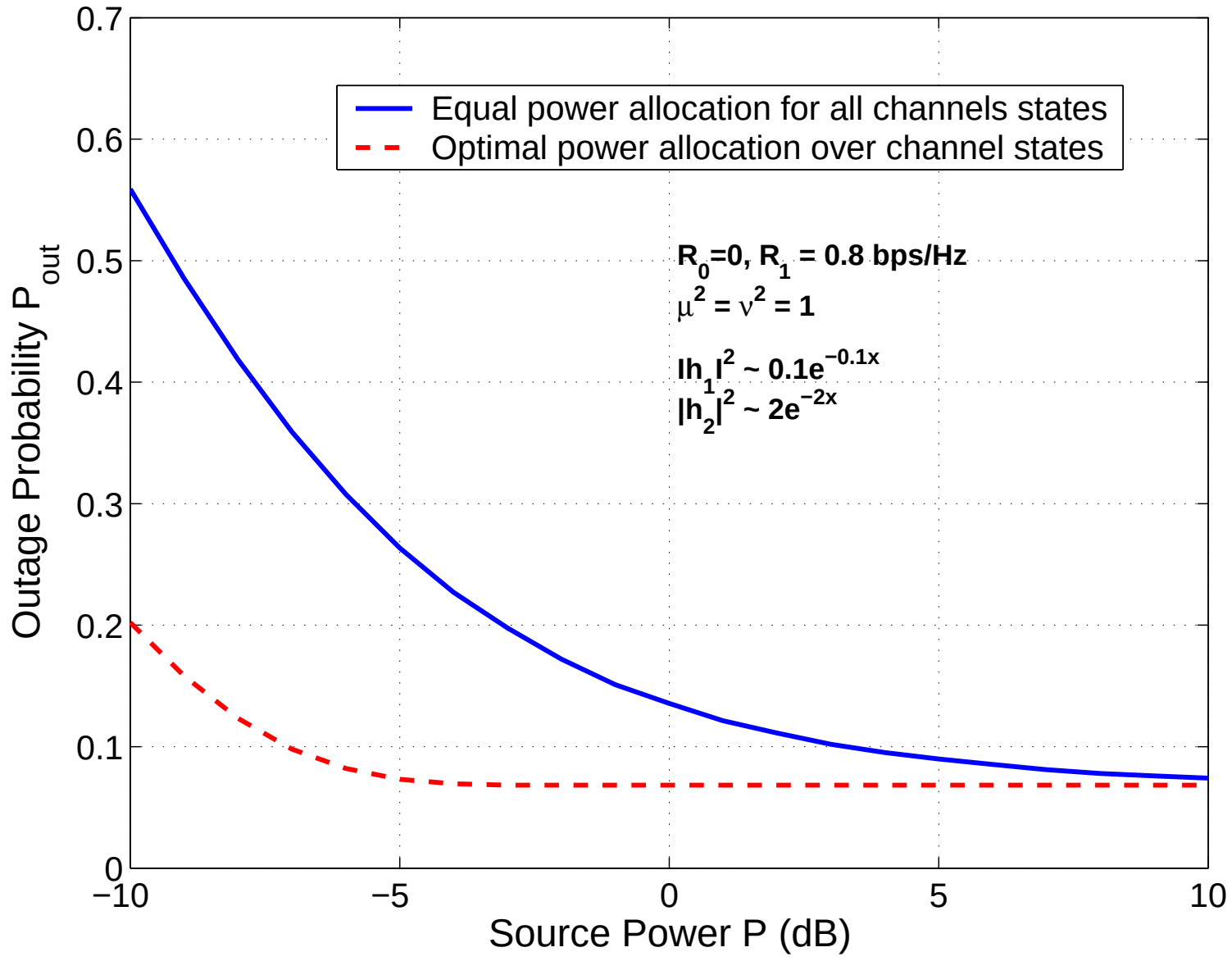

Figure 11: Comparison of outage probabilities achieved by optimal power allocation with those achieved by equal power allocation for a Rayleigh fading BCC with $R_0 = 0$

# 7 Conclusions

We have established the secrecy capacity region of the parallel BCC, where a converse proof is given to show that independent input to each subchannel is optimal. We have also established the secrecy capacity region for the parallel Gaussian BCC, and have characterized the optimal power allocations that achieve the boundary of this region. One fundamental result we have established is the secrecy capacity region of the Gaussian BCC, which complements the secrecy capacity region of the discrete memoryless BCC given by Csiszár and Körner in [5].

We have further applied our results to obtain the ergodic secrecy capacity region for the fading BCC and the optimal power allocations that achieve the boundary of the secrecy capacity region. Our results generalize the secrecy capacity results that have been recently obtained in [12], [13] and [14] (full CSI case). We have also studied the outage performance of the fading BCC. We have obtained the power allocation that minimizes the outage probability that certain target rates are not achieved.

# Appendix

# A Proof of Theorem 1

The achievability follows from [5, Corollary 1] by setting $Q=(Q_1,\ldots,Q_L)$ ($Q$ is indicated by $U$ in [5]), $U=(U_1,\ldots,U_L)$ ($U$ is indicated by $V$ in [5]), $X=(X_1,\ldots,X_L)$, $Y=(Y_1,\ldots,Y_L)$, and $Z=(Z_1,\ldots,Z_L)$ with $Q$, $U$ and $X$ having independent components. Furthermore, we choose the components of these random vectors to satisfy the Markov chain conditions: $Q_l \rightarrow U_l \rightarrow X_l \rightarrow (Y_l, Z_l)$ for $l=1,2,\ldots,L$.

To show the converse, we consider a code $\left(2^{nR_0}, 2^{nR_1}, n\right)$ with the average error probability $P_e$. The probability distribution on $W_0 \times W_1 \times \mathcal{X}^n_{[1,L]} \times \mathcal{Y}^n_{[1,L]} \times \mathcal{Z}^n_{[1,L]}$ is given by

$$\begin{aligned} & p(w_0, w_1, x^n_{[1,L]}, y^n_{[1,L]}, z^n_{[1,L]}) \\ & \quad = p(w_0)p(w_1)p(x^n_{[1,L]}|w_0,w_1)\prod_{i=1}^{n}\prod_{l=1}^{L} p_l(y_{li}, z_{li}|x_{li}) \end{aligned} \tag{61}$$

By Fano's inequality [46, Sec. 2.11], we have

$$\begin{aligned} &H(W_0, W_1|Y^n_{[1,L]}) \leq n(R_0 + R_1)P_e + 1 := n\delta_1 \\ &H(W_0|Z^n_{[1,L]}) \leq nR_0P_e + 1 := n\delta_2 \end{aligned} \tag{62}$$

where $\delta_1, \delta_2 \to 0$ if $P_e \to 0$.

For $l = 1, 2, \ldots, L$, we define the following auxiliary random variables:

$$\begin{aligned} &Q_{li} := (W_0, Y^n_{[1,l-1]}, Y_l^{i-1}, Z^n_{l[i+1]}, Z^n_{[l+1,L]}), \\ &U_{li} := (W_1, Q_{li}). \end{aligned} \tag{63}$$

We note that $(Q_{li}, U_{li}, X_{li}, Y_{li}, Z_{li})$ satisfies the following Markov chain condition:

$$Q_{li} \to U_{li} \to X_{li} \to (Y_{li}, Z_{li}). \tag{64}$$

We first bound the common rate $R_0$ as follows.

$$\begin{aligned} nR_0 = H(W_0) &= I\big(W_0; Y^n_{[1,L]}\big) + H\big(W_0|Y^n_{[1,L]}\big) \\ &\overset{(a)}{\leq} \sum_{l=1}^{L} I(W_0; Y_l^n|Y^n_{[1,l-1]}) + n\delta_1 \\ &\overset{(b)}{=} \sum_{l=1}^{L}\sum_{i=1}^{n} I(W_0; Y_{li}|Y^n_{[1,l-1]}Y_l^{i-1}) + n\delta_1 \\ &\overset{(c)}{\leq} \sum_{l=1}^{L}\sum_{i=1}^{n} I(W_0Y^n_{[1,l-1]}Y_l^{i-1}Z^n_{l[i+1]}Z^n_{[l+1,L]}; Y_{li}) + n\delta_1 \\ &\overset{(d)}{=} \sum_{l=1}^{L}\sum_{i=1}^{n} I(Q_{li}; Y_{li}) + n\delta_1 \end{aligned} \tag{65}$$

where $(a)$ follows from the chain rule and Fano's inequality, $(b)$ follows from the chain rule, $(c)$ follows because $I(A; B|C) \leq I(A, C; B)$, and $(d)$ follows from the definition (63).

We can also bound the common rate $R_0$ as follows.

$$\begin{aligned} nR_0 &= H(W_0) = I\big(W_0; Z_{[1,L]}^n\big) + H\big(W_0|Z_{[1,L]}^n\big) \\ &\leq \sum_{l=1}^{L} I(W_0; Z_l^n|Z_{[l+1,L]}^n) + n\delta_2 \\ &= \sum_{l=1}^{L}\sum_{i=1}^{n} I(W_0; Z_{li}|Z_{l[i+1]}^n Z_{[l+1,L]}^n) + n\delta_2 \\ &\leq \sum_{l=1}^{L}\sum_{i=1}^{n} I(W_0 Y_{[1,l-1]}^n Y_l^{i-1} Z_{l[i+1]}^n Z_{[l+1,L]}^n; Z_{li}) + n\delta_2 \\ &= \sum_{l=1}^{L}\sum_{i=1}^{n} I(Q_{li}; Z_{li}) + n\delta_2 \end{aligned} \tag{66}$$

We now bound the rate $R_1$ and obtain:

$$
\begin{aligned}
nR_1 &\overset{(a)}{=} nR_e \le H\big(W_1|Z^n_{[1,L]}\big) \\
&= H\big(W_1|W_0Z^n_{[1,L]}\big) + I(W_1;W_0|Z^n_{[1,L]}) \\
&\overset{(b)}{\le} I(W_1;Y^n_{[1,L]}|W_0) - I(W_1;Z^n_{[1,L]}|W_0) + H\big(W_1|W_0Y^n_{[1,L]}\big) + H\big(W_0|Z^n_{[1,L]}\big) \\
&\overset{(c)}{\le} I(W_1;Y^n_{[1,L]}|W_0) - I(W_1;Z^n_{[1,L]}|W_0) + n(\delta_1+\delta_2) \\
&= \sum_{l=1}^{L}\Big[I(W_1;Y^n_l|W_0Y^n_{[1,l-1]}) - I(W_1;Z^n_l|W_0Z^n_{[l+1,L]})\Big] + n(\delta_1+\delta_2) \\
&= \sum_{l=1}^{L}\sum_{i=1}^{n}\Big[I(W_1;Y_{li}|W_0Y^n_{[1,l-1]}Y_l^{i-1}) - I(W_1;Z_{li}|W_0Z^n_{l[i+1]}Z^n_{[l+1,L]})\Big] + n(\delta_1+\delta_2) \\
&\overset{(d)}{=} \sum_{l=1}^{L}\sum_{i=1}^{n}\Big[I(W_1Z^n_{l[i+1]}Z^n_{[l+1,L]};Y_{li}|W_0Y^n_{[1,l-1]}Y_l^{i-1}) - I(Z^n_{l[i+1]}Z^n_{[l+1,L]};Y_{li}|W_0W_1Y^n_{[1,l-1]}Y_l^{i-1}) \\
&\qquad - I(W_1Y^n_{[1,l-1]}Y_l^{i-1};Z_{li}|W_0Z^n_{l[i+1]}Z^n_{[l+1,L]}) + I(Y^n_{[1,l-1]}Y_l^{i-1};Z_{li}|W_0W_1Z^n_{l[i+1]}Z^n_{[l+1,L]})\Big] \\
&\qquad + n(\delta_1+\delta_2) \\
&\overset{(e)}{=} \sum_{l=1}^{L}\sum_{i=1}^{n}\Big[I(W_1Z^n_{l[i+1]}Z^n_{[l+1,L]};Y_{li}|W_0Y^n_{[1,l-1]}Y_l^{i-1}) - I(W_1Y^n_{[1,l-1]}Y_l^{i-1};Z_{li}|W_0Z^n_{l[i+1]}Z^n_{[l+1,L]})\Big] \\
&\qquad + n(\delta_1+\delta_2) \\
&= \sum_{l=1}^{L}\sum_{i=1}^{n}\Big[I(Z^n_{l[i+1]}Z^n_{[l+1,L]};Y_{li}|W_0Y^n_{[1,l-1]}Y_l^{i-1}) + I(W_1;Y_{li}|W_0Y^n_{[1,l-1]}Y_l^{i-1}Z^n_{l[i+1]}Z^n_{[l+1,L]}) \\
&\qquad - I(Y^n_{[1,l-1]}Y_l^{i-1};Z_{li}|W_0Z^n_{l[i+1]}Z^n_{[l+1,L]}) - I(W_1;Z_{li}|W_0Y^n_{[1,l-1]}Y_l^{i-1}Z^n_{l[i+1]}Z^n_{[l+1,L]})\Big] \\
&\qquad + n(\delta_1+\delta_2) \\
&\overset{(f)}{=} \sum_{l=1}^{L}\sum_{i=1}^{n}\big[I(W_1;Y_{li}|W_0Y^n_{[1,l-1]}Y_l^{i-1}Z^n_{l[i+1]}Z^n_{[l+1,L]}) - I(W_1;Z_{li}|W_0Y^n_{[1,l-1]}Y_l^{i-1}Z^n_{l[i+1]}Z^n_{[l+1,L]})\Big] \\
&\qquad + n(\delta_1+\delta_2) \\
&\overset{(g)}{=} \sum_{l=1}^{L}\sum_{i=1}^{n}\Big[I(U_{li};Y_{li}|Q_{li}) - I(U_{li};Z_{li}|Q_{li})\Big] + n(\delta_1+\delta_2)
\end{aligned}
\tag{67}
$$

where $(a)$ follows from the perfect secrecy condition, $(b)$ follows because $I(W_1;W_0|Z^n_{[1,L]}) \le H\big(W_0|Z^n_{[1,L]}\big)$, $(c)$ follows from Fano's inequality, $(d)$ follows from the chain rule, $(e)$ and $(f)$ follow from Lemma 7 in [5], and $(g)$ follows from the definition (63).

We introduce a random variable $G$ that is independent of all other random variables, and is uniformly distributed over $\{1, 2, \ldots, n\}$. Define $Q_l = (G, Q_{lG})$, $U_l = (G, U_{lG})$, $X_l = X_{lG}$, $Y_l = Y_{lG}$, and $Z_l = Z_{lG}$ for $l = 1, \ldots, L$. Note that $(Q_l, U_l, X_l, Y_l, Z_l)$ satisfies the following Markov chain condition:

$$Q_l \to U_l \to X_l \to (Y_l, Z_l), \qquad \text{for } \ l = 1, \ldots, L. \tag{68}$$

Using the above definitions, (65), (66) and (67) become

$$\begin{aligned}
R_0 &\leq \sum_{l=1}^{L} I(Q_{lG}; Y_{lG}|G) + \delta_1 \leq \sum_{l=1}^{L} I(Q_l; Y_l) + \delta_1 \\
R_0 &\leq \sum_{l=1}^{L} I(Q_{lG}; Z_{lG}|G) + \delta_2 \leq \sum_{l=1}^{L} I(Q_l; Z_l) + \delta_2 \\
R_1 &\leq \sum_{l=1}^{L} I(U_l; Y_l|Q_l) - I(U_l; Z_l|Q_l)\Big] + \delta_1 + \delta_2
\end{aligned} \tag{69}$$

Therefore, an outer bound on the secrecy capacity region $\mathcal{C}_s$ is given by the following set:

$$\bigcup \Big\{ (R_0, R_1) \text{ that satisfy (69)} \Big\} \tag{70}$$

where the union is over all probability distributions $p(q_{[1,L]}, u_{[1,L]}, x_{[1,L]}, y_{[1,L]}, z_{[1,L]})$. Finally, we note that each term in (69) depends only on the distribution $p(q_l, u_l, x_l, y_l, z_l)$. Hence there is no loss of optimality to consider only those distributions that have the form $\prod_{l=1}^{L} p(q_l, u_l, x_l) p(y_l, z_l|x_l)$. This concludes the converse proof.

# B Proof of Theorem 2

By Lemma 1, we need to prove Theorem 2 only for the channel defined by (24).

The achievability follows by applying Corollary 3 and choosing the following input distribution

$$\begin{aligned}
&\text{for } l \in A, \quad Q_l \sim \mathcal{N}(0, p_{l0}), \quad X'_l \sim \mathcal{N}(0, p_{l1}) \quad \text{with } X'_l \text{ independent of } Q_l, \\
&\qquad\qquad\quad X_l = Q_l + X'_l; \\
&\text{for } l \in A^c; \quad X_l \sim \mathcal{N}(0, p_{l0}).
\end{aligned} \tag{71}$$

To show the converse, we first apply (65) and obtain

$$
\begin{aligned}
nR_0 &\le \sum_{l=1}^{L}\sum_{i=1}^{n} I(Q_{li};Y_{li}) + n\delta_1 \\
&\le \sum_{l\in A}\sum_{i=1}^{n} I(Q_{li};Y_{li}) + \sum_{l\in A^c}\sum_{i=1}^{n} I(X_{li};Y_{li}) + n\delta_1 \\
&= \sum_{l\in A}\sum_{i=1}^{n}[h(Y_{li}) - h(Y_{li}|Q_{li})] + \sum_{l\in A^c}\sum_{i=1}^{n}[h(Y_{li}) - h(Y_{li}|X_{li})] + n\delta_1 \\
&\le \sum_{l\in A}\sum_{i=1}^{n}\frac{1}{2}\log 2\pi e\left(EX_{li}^2+\mu_l^2\right) - \sum_{l\in A}\sum_{i=1}^{n} h(Y_{li}|Q_{li}) \\
&\quad + \sum_{l\in A^c}\sum_{i=1}^{n}\frac{1}{2}\log 2\pi e\left(EX_{li}^2+\mu_l^2\right) - \sum_{l\in A^c}\frac{n}{2}\log 2\pi e\mu_l^2 + n\delta_1 \\
&\le \sum_{l\in A}\frac{n}{2}\log 2\pi e\left(\frac{1}{n}\sum_{i=1}^{n} EX_{li}^2+\mu_l^2\right) - \sum_{l\in A}\sum_{i=1}^{n} h(Y_{li}|Q_{li}) \\
&\quad + \sum_{l\in A^c}\frac{n}{2}\log 2\pi e\left(\frac{1}{n}\sum_{i=1}^{n} EX_{li}^2+\mu_l^2\right) - \sum_{l\in A^c}\frac{n}{2}\log 2\pi e\mu_l^2 + n\delta_1 \\
&\le \sum_{l\in A}\frac{n}{2}\log 2\pi e\left(p_l+\mu_l^2\right) - \sum_{l\in A}\sum_{i=1}^{n} h(Y_{li}|Q_{li}) + \sum_{l\in A^c}\frac{n}{2}\log\left(1+\frac{p_l}{\mu_l^2}\right) + n\delta_1
\end{aligned} \tag{72}
$$

where we define $p_l = \frac{1}{n}\sum_{i=1}^{n} EX_{li}^2$.

It is easy to see that for $l \in A$

$$
\sum_{i=1}^{n} h(Y_{li}|Q_{li}) \le \sum_{i=1}^{n} h(Y_{li}) \le \frac{n}{2}\log 2\pi e\left(p_l+\mu_l^2\right) \tag{73}
$$

and

$$
\sum_{i=1}^{n} h(Y_{li}|Q_{li}) \ge \sum_{i=1}^{n} h(Y_{li}|X_{li}) = \frac{n}{2}\log 2\pi e\mu_l^2 \tag{74}
$$

Hence there exists $0 \le \beta_l \le 1$ such that

$$
\sum_{i=1}^{n} h(Y_{li}|Q_{li}) = \frac{n}{2}\log 2\pi e(\beta_l p_l + \mu_l^2) \tag{75}
$$

Applying (75) to (72), we obtain

$$
nR_0 \le \sum_{l\in A}\frac{n}{2}\log\left(1+\frac{(1-\beta_l)p_l}{\beta_l p_l+\mu_l^2}\right) + \sum_{l\in A^c}\frac{n}{2}\log\left(1+\frac{p_l}{\mu_l^2}\right) + n\delta_1 \tag{76}
$$

We apply (66), follow the steps that are similar to those in (72), and obtain the following bound

$$nR_0 \le \sum_{l\in A} \frac{n}{2}\log 2\pi e\left(p_l+\nu_l^2\right) - \sum_{l\in A}\sum_{i=1}^{n} h(Z_{li}|Q_{li}) + \sum_{l\in A^c}\frac{n}{2}\log\left(1+\frac{p_l}{\nu_l^2}\right) + n\delta_2 \tag{77}$$

For the second term in the preceding equation, we apply the entropy power inequality and obtain

$$\begin{aligned} h(Z_{li}|Q_{li}=q_{li}) &= h(Y_{li}+V'_{li}|Q_{li}=q_{li}) \\ &\ge \frac{1}{2}\log\left(2^{2h(Y_{li}|Q_{li}=q_{li})}+2^{2h(V'_{li}|Q_{li}=q_{li})}\right) \\ &= \frac{1}{2}\log\left(2^{2h(Y_{li}|Q_{li}=q_{li})}+2\pi e(\nu_l^2-\mu_l^2)\right) \end{aligned} \tag{78}$$

Hence

$$\begin{aligned} \sum_{i=1}^{n} h(Z_{li}|Q_{li}) &\ge \frac{1}{2}\sum_{i=1}^{n} \mathrm{E}\log\left(2^{2h(Y_{li}|Q_{li}=q_{li})}+2\pi e(\nu_l^2-\mu_l^2)\right) \\ &\overset{(a)}{\ge} \frac{1}{2}\sum_{i=1}^{n}\log\left(2^{2\mathrm{E}h(Y_{li}|Q_{li}=q_{li})}+2\pi e(\nu_l^2-\mu_l^2)\right) \\ &= \frac{1}{2}\sum_{i=1}^{n}\log\left(2^{2h(Y_{li}|Q_{li})}+2\pi e(\nu_l^2-\mu_l^2)\right) \\ &\overset{(b)}{\ge} \frac{n}{2}\log\left(2^{2\frac{1}{n}\sum_{i=1}^{n}h(Y_{li}|Q_{li})}+2\pi e(\nu_l^2-\mu_l^2)\right) \\ &\overset{(c)}{=} \frac{n}{2}\log\left(2\pi e(\beta_l p_l+\mu_l^2)+2\pi e(\nu_l^2-\mu_l^2)\right) \\ &= \frac{n}{2}\log\left(2\pi e(\beta_l p_l+\nu_l^2)\right) \end{aligned} \tag{79}$$

where $(a)$ and $(b)$ follow from Jensen's inequality and the fact that $\log(2^x+c)$ is a convex function of $x$, and $(c)$ follows from (75).

By applying (79) to (77), we obtain

$$nR_0 \le \sum_{l\in A}\frac{n}{2}\log\left(1+\frac{(1-\beta_l)p_l}{\beta_l p_l+\nu_l^2}\right) + \sum_{l\in A^c}\frac{n}{2}\log\left(1+\frac{p_l}{\nu_l^2}\right) + n\delta_2 \tag{80}$$

We apply (67), follow the steps that are similar to those in (21), and obtain the following

bound

$$
\begin{aligned}
nR_1 &\leq \sum_{l\in A}\sum_{i=1}^{n}\Big[I(X_{li};Y_{li}|Q_{li}) - I(X_{li};Z_{li}|Q_{li})\Big] + n(\delta_1+\delta_2) \\
&= \sum_{l\in A}\sum_{i=1}^{n}\Big[h(Y_{li}|Q_{li}) - h(Y_{li}|X_{li};Q_{li}) - h(Z_{li}|Q_{li}) + h(Z_{li}|X_{li};Q_{li})\Big] + n(\delta_1+\delta_2) \quad (81)\\
&\leq \sum_{l\in A}\left[\frac{n}{2}\log\left(1+\frac{\beta_l p_l}{\mu_l^2}\right) - \frac{n}{2}\log\left(1+\frac{\beta_l p_l}{\nu_l^2}\right)\right] + n(\delta_1+\delta_2)
\end{aligned}
$$

where the last equality follows from (75) and (79).

For $l \in A$, we define $p_{l0} = (1-\beta_l)p_l$ and $p_{l1} = \beta_l p_l$. For $l \in A^c$, we define $p_{l0} = p_l$. It is clear from (23) that

$$\sum_{l\in A}[p_{l0}+p_{l1}] + \sum_{l\in A^c} p_{l0} \leq P \quad (82)$$

This concludes the proof of the converse.

# C Proof of Theorem 3

We apply Lemma 2 and consider the following three cases. For each case, we apply the technique in [8] to solve the optimization problem.

**Case 1:** We need to find $\underline{p}^{(1)} \in \mathcal{P}$ that maximizes $\gamma_0 R_{01}(\underline{p}) + \gamma_1 R_1(\underline{p})$. If $\underline{p}^{(1)}$ satisfies $R_{01}\left(\underline{p}^{(1)}\right) < R_{02}\left(\underline{p}^{(1)}\right)$, then the optimal $\underline{p}^* = \underline{p}^{(1)}$.

The Lagrangian is given by

$$
\begin{aligned}
\mathcal{L} =& \sum_{l\in A}\left[\frac{\gamma_0}{2}\log\left(1+\frac{p_{l0}}{\mu_l^2+p_{l1}}\right) + \frac{\gamma_1}{2}\log\left(1+\frac{p_{l1}}{\mu_l^2}\right) - \frac{\gamma_1}{2}\log\left(1+\frac{p_{l1}}{\nu_l^2}\right)\right] \\
&+ \sum_{l\in A^c}\frac{\gamma_0}{2}\log\left(1+\frac{p_{l0}}{\mu_l^2}\right) - \lambda\left[\sum_{l\in A}[p_{l0}+p_{l1}] + \sum_{l\in A^c}p_{l0}\right]
\end{aligned} \quad (83)
$$

where $\lambda$ is a Lagrange multiplier.

For $l \in A^c$, $p_{l0}^{(1)}$ needs to maximize the following $\mathcal{L}_l$

$$\mathcal{L}_l = \frac{\gamma_0}{2}\log\left(1+\frac{p_{l0}}{\mu_l^2}\right) - \lambda\, p_{l0} = \int_0^{p_{l0}}\left(\frac{\gamma_0}{2\ln 2}\frac{1}{\mu_l^2+x} - \lambda\right)dx \quad (84)$$

It is clear that $p_{l0}^{(1)}$ that optimizes $\mathcal{L}_l$ is either the root of the following equation

$$\frac{\gamma_0}{2\ln 2}\frac{1}{\mu_l^2+x} - \lambda = 0 \quad (85)$$

if the root is positive or zero, i.e.,

$$p_{l0}^{(1)} = \left(\frac{\gamma_0}{2\lambda \ln 2} - \mu_l^2\right)^+. \tag{86}$$

For $l \in A$, $p_{l0}^{(1)}$ and $p_{l1}^{(1)}$ need to maximize the following $\mathcal{L}_l$

$$\begin{aligned} \mathcal{L}_l &= \frac{\gamma_0}{2}\log\left(1+\frac{p_{l0}}{\mu_l^2+p_{l1}}\right)+\frac{\gamma_1}{2}\log\left(1+\frac{p_{l1}}{\mu_l^2}\right)-\frac{\gamma_1}{2}\log\left(1+\frac{p_{l1}}{\nu_l^2}\right)-\lambda(p_{l0}+p_{l1}) \\ &= \int_{p_{l1}}^{p_{l1}+p_{l0}} u_{l0}^{(1)}(x)dx + \int_0^{p_{l1}} u_{l1}^{(1)}(x)dx \\ &\le \int_0^{\infty}\left(\max\left\{u_{l0}^{(1)}(x), u_{l1}^{(1)}(x)\right\}\right)^+ dx \end{aligned} \tag{87}$$

where

$$\begin{aligned} u_{l0}^{(1)}(x) &= \frac{\gamma_0}{2\ln 2}\frac{1}{\mu_l^2+x}-\lambda \\ u_{l1}^{(1)}(x) &= \frac{\gamma_1}{2\ln 2}\left(\frac{1}{\mu_l^2+x}-\frac{1}{\nu_l^2+x}\right)-\lambda \end{aligned} \tag{88}$$

We next derive $p_{l0}^{(1)}$ and $p_{l0}^{(1)}$ that achieve the upper bound on $\mathcal{L}_l$ in (87) and hence maximize $\mathcal{L}_l$.

We define $x_{l0}^{(1)}$ to be the root of $u_{l0}^{(1)}(x) = 0$ and $x_{l1}^{(1)}$ to be the largest root of $u_{l1}^{(1)}(x) = 0$, i.e.,

$$\begin{aligned} x_{l0}^{(1)} &= \frac{\gamma_0}{2\lambda\ln 2} - \mu_l^2 \\ x_{l1}^{(1)} &= \frac{1}{2}\sqrt{(\nu_l^2-\mu_l^2)\left(\nu_l^2-\mu_l^2+\frac{2\gamma_1}{\lambda\ln 2}\right)} - \frac{1}{2}(\mu_l^2+\nu_l^2). \end{aligned} \tag{89}$$

It can be seen that $u_{l0}^{(1)}(x)$ and $u_{l1}^{(1)}(x)$ intersect only once at

$$x_{lr}^{(1)} = \frac{\gamma_1}{\gamma_0}(\nu_l^2-\mu_l^2) - \nu_l^2. \tag{90}$$

In the following, we consider two cases.

**(1)** $\frac{\gamma_1}{\gamma_0} > \frac{\nu_l^2}{\nu_l^2-\mu_l^2}$, i.e., $x_{lr}^{(1)}$ is positive.

It is easy to see that $u_{l1}^{(1)}(0) > u_{l0}^{(1)}(0)$. The optimal $p_{l0}^{(1)}$ and $p_{l1}^{(1)}$ depend on the value of $\lambda$ and fall into the following three possibilities.

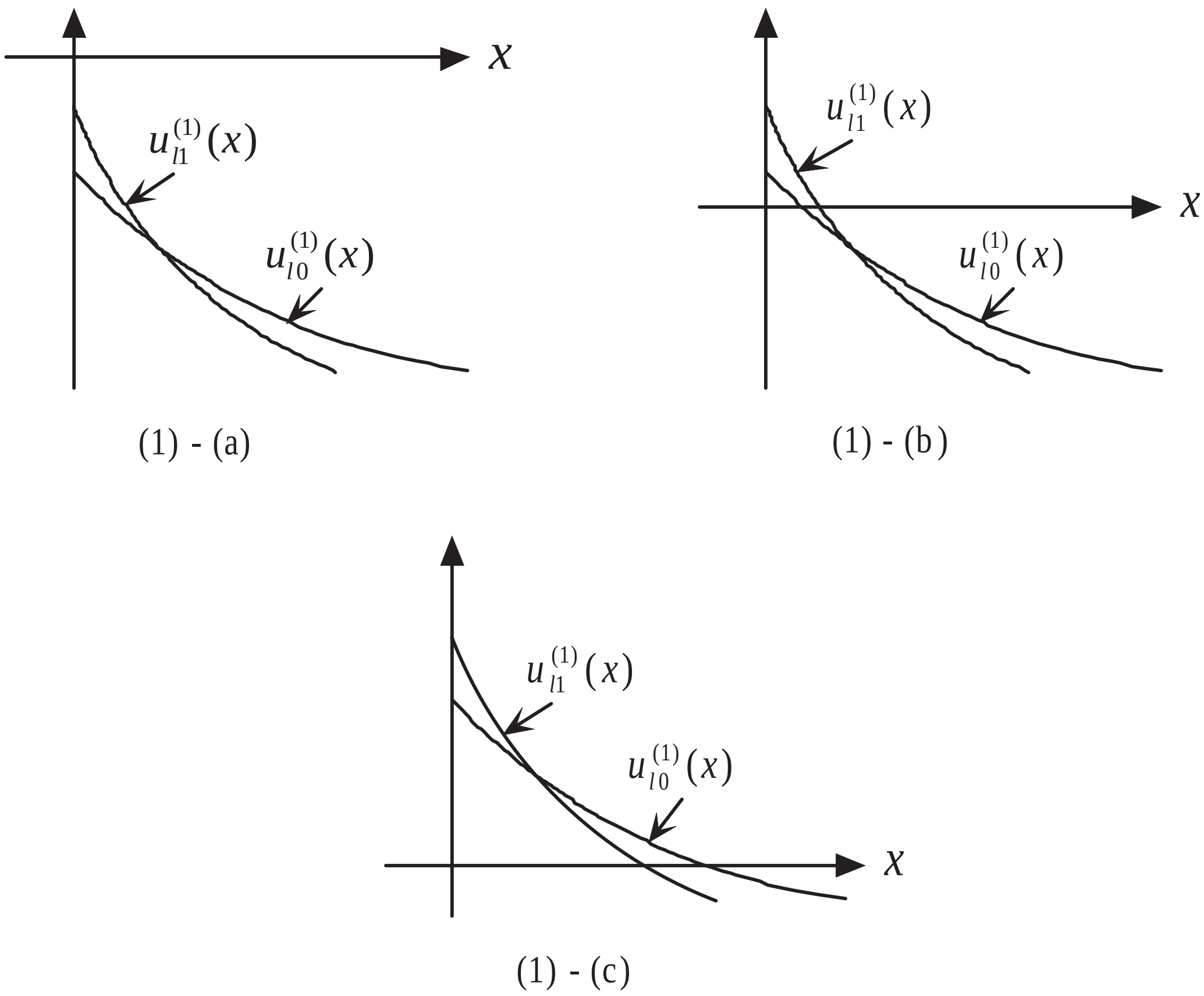

Figure 12: Illustration of $u_{l0}^{(1)}(x)$ and $u_{l1}^{(1)}(x)$ for Case 1 with $\frac{\gamma_1}{\gamma_0} > \frac{\nu_l^2}{\nu_l^2-\mu_l^2}$

(a) If $u_{l1}^{(1)}(0) < 0$ (see Fig. 12 (1)-(a)), i.e., $\frac{\gamma_1(\nu_l^2-\mu_l^2)}{2\mu_l^2\nu_l^2 \ln 2} < \lambda$, then both $u_{l0}^{(1)}(x)$ and $u_{l1}^{(1)}(x)$ are negative for $x > 0$. The upper bound on $\mathcal{L}_l$ in (87) is achieved by $p_{l0}^{(1)} = 0$ and $p_{l1}^{(1)} = 0$.

(b) If $u_{l1}^{(1)}(0) \geq 0$ and $x_{l0}^{(1)} < x_{lr}^{(1)}$ (see Fig. 12 (1)-(b)), i.e., $\frac{\gamma_0^2}{2\ln 2(\gamma_1-\gamma_0)(\nu_l^2-\mu_l^2)} < \lambda \leq \frac{\gamma_1(\nu_l^2-\mu_l^2)}{2\mu_l^2\nu_l^2 \ln 2}$, then the upper bound on $\mathcal{L}_l$ in (87) is achieved by $p_{l0}^{(1)} = 0$ and $p_{l1}^{(1)} = x_{l1}^{(1)}$.

(c) If $x_{l0}^{(1)} \geq x_{lr}^{(1)}$ (see Fig. 12 (1)-(c)), i.e., $\lambda \leq \frac{\gamma_0^2}{2\ln 2(\gamma_1-\gamma_0)(\nu_l^2-\mu_l^2)}$, then the upper bound on $\mathcal{L}_l$ in (87) is achieved by $p_{l0}^{(1)} = x_{l0}^{(1)} - x_{lr}^{(1)}$ and $p_{l1}^{(1)} = x_{lr}^{(1)}$.

In summary, we obtain

$$p_{l0}^{(1)} = \left[x_{l0}^{(1)} - x_{lr}^{(1)}\right]^+, \qquad p_{l1}^{(1)} = \left[\min\left\{x_{l1}^{(1)}, x_{lr}^{(1)}\right\}\right]^+. \tag{91}$$

**(2)** $\frac{\gamma_1}{\gamma_0} \leq \frac{\nu_l^2}{\nu_l^2-\mu_l^2}$, i.e., $x_{lr}^{(1)}$ is zero or negative.

It is easy to see that $u_0^{(1)}(0) \geq u_1^{(1)}(0)$.

(a) If $u_{l0}^{(1)}(0) \leq 0$, i.e., $\frac{\gamma_0}{2\mu_l^2 \ln 2} \leq \lambda$ (see Fig. 13 (2)-(a)), then the upper bound on $\mathcal{L}_l$ in (87) is achieved by $p_{l0}^{(1)} = 0$ and $p_{l1}^{(1)} = 0$.

(b) If $u_{l0}^{(1)}(0) > 0$, i.e., $\lambda < \frac{\gamma_0}{2\mu_l^2 \ln 2}$ (see Fig. 13 (2)-(b)), the upper bound on $\mathcal{L}_l$ in (87) is achieved by $p_{l0}^{(1)} = x_{l0}^{(1)}$ and $p_{l1}^{(1)} = 0$.

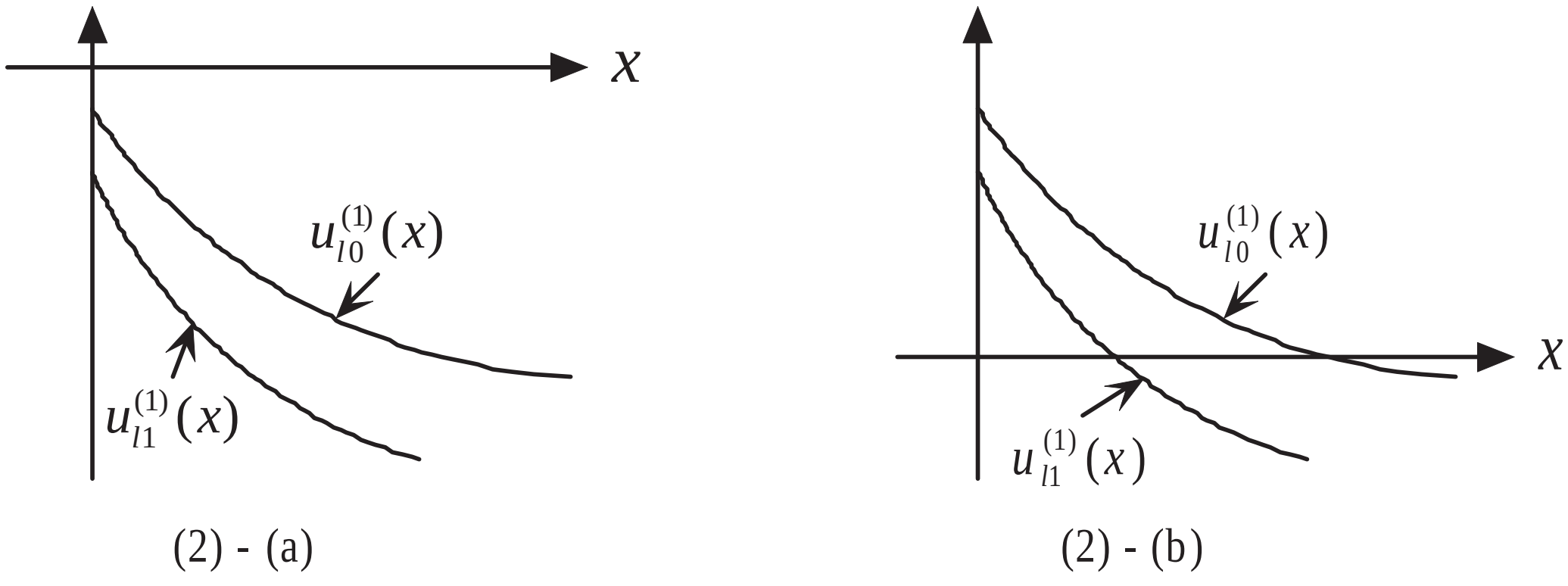

Figure 13: Illustration of $u_{l0}^{(1)}(x)$ and $u_{l1}^{(1)}(x)$ for Case 1 with $\frac{\gamma_1}{\gamma_0} \leq \frac{\nu_l^2}{\nu_l^2 - \mu_l^2}$

In summary, we obtain

$$p_{l0}^{(1)} = \left[x_{l0}^{(1)}\right]^+, \qquad p_{l1}^{(1)} = 0. \tag{92}$$

The Lagrange parameter $\lambda$ needs to be chosen to satisfy the power constraint

$$\sum_{l \in A} [p_{l0} + p_{l1}] + \sum_{l \in A^c} p_{l0} \leq P. \tag{93}$$

According to Lemma 2, if the condition $R_{01}\left(\underline{p}^{(1)}\right) < R_{02}\left(\underline{p}^{(1)}\right)$ is satisfied, then the optimal $\underline{p}^* = \underline{p}^{(1)}$.

**Case 2:** We need to find $\underline{p}^{(2)} \in \mathcal{P}$ that maximizes $\gamma_0 R_{02}(\underline{p}) + \gamma_1 R_1(\underline{p})$. If $\underline{p}^{(2)}$ satisfies $R_{01}\left(\underline{p}^{(2)}\right) > R_{02}\left(\underline{p}^{(2)}\right)$, then the optimal $\underline{p}^* = \underline{p}^{(2)}$.

The Lagrangian is given by

$$\begin{aligned}\mathcal{L} =& \sum_{l \in A} \left[\frac{\gamma_0}{2} \log\left(1 + \frac{p_{l0}}{\nu_l^2 + p_{l1}}\right) + \frac{\gamma_1}{2} \log\left(1 + \frac{p_{l1}}{\mu_l^2}\right) - \frac{\gamma_1}{2} \log\left(1 + \frac{p_{l1}}{\nu_l^2}\right)\right] \\ &+ \sum_{l \in A^c} \frac{\gamma_0}{2} \log\left(1 + \frac{p_{l0}}{\nu_l^2}\right) - \lambda \left[\sum_{l \in A} [p_{l0} + p_{l1}] + \sum_{l \in A^c} p_{l0}\right]\end{aligned} \tag{94}$$

where $\lambda$ is a Lagrange multiplier.

For $l \in A^c$, it is easy to see that

$$p_{l0}^{(2)} = \left(\frac{\gamma_0}{2\lambda \ln 2} - \nu_l^2\right)^+. \tag{95}$$

For $l \in A$, $p_{l0}^{(2)}$ and $p_{l1}^{(2)}$ need to maximize the following $\mathcal{L}_l$

$$\begin{aligned}
\mathcal{L}_l &= \frac{\gamma_0}{2}\log\left(1+\frac{p_{l0}}{\nu_l^2+p_{l1}}\right)+\frac{\gamma_1}{2}\log\left(1+\frac{p_{l1}}{\mu_l^2}\right)-\frac{\gamma_1}{2}\log\left(1+\frac{p_{l1}}{\nu_l^2}\right)-\lambda(p_{l0}+p_{l1})\\
&= \int_{p_{l1}}^{p_{l1}+p_{l0}} u_{l0}^{(2)}(x)dx + \int_0^{p_{l1}} u_{l1}^{(2)}(x)dx \\
&\leq \int_0^{\infty}\left(\max\left\{u_{l0}^{(2)}(x), u_{l1}^{(2)}(x)\right\}\right)^+ dx
\end{aligned} \tag{96}$$

where

$$\begin{aligned}
u_{l0}^{(2)}(x) &= \frac{\gamma_0}{2\ln 2}\frac{1}{\nu_l^2+x}-\lambda\\
u_{l1}^{(2)}(x) &= \frac{\gamma_1}{2\ln 2}\left(\frac{1}{\mu_l^2+x}-\frac{1}{\nu_l^2+x}\right)-\lambda
\end{aligned} \tag{97}$$

We define $x_{l0}^{(2)}$ to be the root of $u_{l0}^{(2)}(x)=0$ and $x_{l1}^{(2)}$ to be the root of $u_{l1}^{(2)}(x)=0$, i.e.,

$$\begin{aligned}
x_{l0}^{(2)} &= \frac{\gamma_0}{2\lambda\ln 2}-\nu_l^2\\
x_{l1}^{(2)} &= \frac{1}{2}\sqrt{(\nu_l^2-\mu_l^2)\left(\nu_l^2-\mu_l^2+\frac{2\gamma_1}{\lambda\ln 2}\right)}-\frac{1}{2}(\mu_l^2+\nu_l^2)
\end{aligned} \tag{98}$$

It is clear that $u_{l0}^{(2)}(x)$ and $u_{l1}^{(2)}(x)$ intersect only once at

$$x_{lr}^{(2)} = \frac{\gamma_1}{\gamma_0}(\nu_l^2-\mu_l^2)-\mu_l^2 \tag{99}$$

Following steps similar to those in Case 1, we obtain the following $p_{l0}^{(2)}$ and $p_{l0}^{(2)}$ that achieve the upper bound on $\mathcal{L}_l$ in (96) and hence maximize $\mathcal{L}_l$.

(1) If $\frac{\gamma_1}{\gamma_0} > \frac{\mu_l^2}{\nu_l^2-\mu_l^2}$, i.e., $x_{lr}^{(2)}$ is positive,

$$p_{l0}^{(2)} = \left[x_{l0}^{(2)} - x_{lr}^{(2)}\right]^+, \qquad p_{l1}^{(2)} = \left[\min\left\{x_{l1}^{(2)}, x_{lr}^{(2)}\right\}\right]^+. \tag{100}$$

(2) If $\frac{\gamma_1}{\gamma_0} \leq \frac{\mu_l^2}{\nu_l^2-\mu_l^2}$, i.e., $x_{lr}^{(2)}$ is zero or negative,

$$p_{l0}^{(2)} = \left[x_{l0}^{(2)}\right]^+, \qquad p_{l1}^{(2)} = 0. \tag{101}$$

The Lagrange parameter $\lambda$ needs to be chosen to satisfy the power constraint. According to Lemma 2, if the condition $R_{01}\left(\underline{p}^{(2)}\right) > R_{02}\left(\underline{p}^{(2)}\right)$ is satisfied, then the optimal $\underline{p}^* = \underline{p}^{(2)}$.

**Case 3:** We need to find $\underline{p}^{(\alpha)} \in \mathcal{P}$ that maximizes $\gamma_0\left(\alpha R_{01}(\underline{p}) + \bar{\alpha} R_{02}(\underline{p})\right) + \gamma_1 R_1(\underline{p})$ for a given $0 \le \alpha \le 1$. We then choose $\alpha$ to satisfy $R_{01}(\underline{p}^{(\alpha)}) = R_{02}(\underline{p}^{(\alpha)})$. According to Lemma 2, if we have not found the optimal $\underline{p}^*$ in Cases 1 and 2, such an $\alpha$ must exist.

The Lagrangian is given by

$$\begin{aligned}\mathcal{L} =& \sum_{l\in A}\left[\frac{\gamma_0\alpha}{2}\log\left(1+\frac{p_{l0}}{\mu_l^2+p_{l1}}\right)+\frac{\gamma_0\bar{\alpha}}{2}\log\left(1+\frac{p_{l0}}{\nu_l^2+p_{l1}}\right)\right.\\ &\left.\qquad+\frac{\gamma_1}{2}\log\left(1+\frac{p_{l1}}{\mu_l^2}\right)-\frac{\gamma_1}{2}\log\left(1+\frac{p_{l1}}{\nu_l^2}\right)\right]\\ &+\sum_{l\in A^c}\left[\frac{\gamma_0\alpha}{2}\log\left(1+\frac{p_{l0}}{\mu_l^2}\right)+\frac{\gamma_0\bar{\alpha}}{2}\log\left(1+\frac{p_{l0}}{\nu_l^2}\right)\right]-\lambda\left[\sum_{l\in A}[p_{l0}+p_{l1}]+\sum_{l\in A^c}p_{l0}\right]\end{aligned} \tag{102}$$

where $\lambda$ is a Lagrange multiplier.

For $l \in A^c$, it is clear that $p_{l0}^{(\alpha)}$ is either the root of the following equation

$$\frac{\gamma_0\alpha}{2\ln 2}\frac{1}{\mu_l^2+x}+\frac{\gamma_0\bar{\alpha}}{2\ln 2}\frac{1}{\nu_l^2+x}=\lambda \tag{103}$$

if the root is positive, or zero. Hence $p_{l0}^{(\alpha)}$ is given by

$$p_{l0}^{(\alpha)} = \left(\frac{1}{2}\sqrt{\left(\nu_l^2-\mu_l^2-\frac{\gamma_0}{2\ln 2\lambda}\right)^2+\frac{2\alpha\gamma_0}{\lambda\ln 2}(\nu_l^2-\mu_l^2)}-\frac{1}{2}\left(\mu_l^2+\nu_l^2-\frac{\gamma_0}{2\ln 2\lambda}\right)\right)^+ \tag{104}$$

For $l \in A$, $p_{l0}^{(\alpha)}$ and $p_{l1}^{(\alpha)}$ need to maximize the following $\mathcal{L}_l$

$$\begin{aligned}\mathcal{L}_l &= \frac{\gamma_0\alpha}{2}\log\left(1+\frac{p_{l0}}{\mu_l^2+p_{l1}}\right)+\frac{\gamma_0\bar{\alpha}}{2}\log\left(1+\frac{p_{l0}}{\nu_l^2+p_{l1}}\right)+\frac{\gamma_1}{2}\log\left(1+\frac{p_{l1}}{\mu_l^2}\right)\\ &\quad-\frac{\gamma_1}{2}\log\left(1+\frac{p_{l1}}{\nu_l^2}\right)-\lambda(p_{l0}+p_{l1})\\ &=\int_{p_{l1}}^{p_{l1}+p_{l0}}u_{l0}^{(\alpha)}(x)dx+\int_0^{p_{l1}}u_{l1}^{(\alpha)}(x)dx\\ &\le\int_0^{\infty}\left(\max\left\{u_{l0}^{(\alpha)}(x),u_{l1}^{(\alpha)}(x)\right\}\right)^+dx\end{aligned} \tag{105}$$

where

$$\begin{aligned}u_{l0}^{(\alpha)}(x) &= \frac{\gamma_0}{2\ln 2}\left(\frac{\alpha}{\mu_l^2+x}+\frac{\bar{\alpha}}{\nu_l^2+x}\right)-\lambda\\ u_{l1}^{(\alpha)}(x) &= \frac{\gamma_1}{2\ln 2}\left(\frac{1}{\mu_l^2+x}-\frac{1}{\nu_l^2+x}\right)-\lambda\end{aligned} \tag{106}$$

We define $x_{l0}^{(\alpha)}$ to be the largest root of $u_{l0}^{(\alpha)}(x) = 0$ and $x_{l1}^{(\alpha)}$ to be the largest root of $u_{l1}^{(\alpha)}(x) = 0$, i.e.,

$$\begin{aligned} x_{l0}^{(\alpha)} &= \frac{1}{2}\sqrt{\left(\nu_l^2 - \mu_l^2 - \frac{\gamma_0}{2\ln 2\lambda}\right)^2 + \frac{2\alpha\gamma_0}{\lambda \ln 2}(\nu_l^2 - \mu_l^2)} - \frac{1}{2}\left(\mu_l^2 + \nu_l^2 - \frac{\gamma_0}{2\ln 2\lambda}\right) \\ x_{l1}^{(\alpha)} &= \frac{1}{2}\sqrt{(\nu_l^2 - \mu_l^2)\left(\nu_l^2 - \mu_l^2 + \frac{2\gamma_1}{\lambda \ln 2}\right)} - \frac{1}{2}(\mu_l^2 + \nu_l^2) \end{aligned} \tag{107}$$

It is clear that $u_{l0}^{(\alpha)}(x)$ and $u_{l1}^{(\alpha)}(x)$ intersect only once at

$$x_{lr}^{(\alpha)} = \frac{\gamma_1}{\gamma_0}(\nu_l^2 - \mu_l^2) - (\alpha\nu_l^2 + \bar{\alpha}\mu_l^2) \tag{108}$$

Following the similar steps to those in Case 1, we obtain

(1) If $\frac{\gamma_1}{\gamma_0} > \frac{\alpha\nu_l^2 + \bar{\alpha}\mu_l^2}{\nu_l^2 - \mu_l^2}$, i.e., $x_{lr}^{(\alpha)}$ is positive,

$$p_{l0}^{(\alpha)} = \left[x_{l0}^{(\alpha)} - x_{lr}^{(\alpha)}\right]^+, \qquad p_{l1}^{(\alpha)} = \left[\min\left\{x_{l1}^{(\alpha)}, x_{lr}^{(\alpha)}\right\}\right]^+. \tag{109}$$

(2) If $\frac{\gamma_1}{\gamma_0} \leq \frac{\alpha\nu_l^2 + \bar{\alpha}\mu_l^2}{\nu_l^2 - \mu_l^2}$, i.e., $x_{lr}^{(\alpha)}$ is negative or zero,

$$p_{l0}^{(\alpha)} = \left[x_{l0}^{(\alpha)}\right]^+, \qquad p_{l1}^{(\alpha)} = 0. \tag{110}$$

The Lagrange parameter $\lambda$ needs to be chosen to satisfy the power constraint. We finally choose $\alpha$ to satisfy $R_{01}\left(\underline{p}^{(\alpha)}\right) = R_{02}\left(\underline{p}^{(\alpha)}\right)$. Then $\underline{p}^* = \underline{p}^{(\alpha)}$.

# References

[1] A. D. Wyner. The wire-tap channel. *Bell Syst. Tech. J.*, 54(8):1355–1387, October 1975.

[2] S. K. Leung-Yan-Cheong and M. E. Hellman. The Gaussian wire-tap channel. *IEEE Trans. Inform. Theory*, 24(4):451–456, July 1978.

[3] P. Parada and R. Blahut. Secrecy capacity of SIMO and slow fading channels. In *Proc. IEEE Int. Symp. Information Theory (ISIT)*, pages 2152–2155, Adelaide, Australia, September 2005.

[4] J. Barros and M. R. D. Rodrigues. Secrecy capacity of wireless channels. In *Proc. IEEE Int. Symp. Information Theory (ISIT)*, Seattle, WA, USA, July 2006.

[5] I. Csiszár and J. Körner. Broadcast channels with confidential messages. *IEEE Trans. Inform. Theory*, 24(3):339–348, May 1978.

[6] R. Liu, I. Maric, P. Spasojevic, and R. Yates. Discrete memoryless interference and broadcast channels with confidential messages. In *Proc. Annu. Allerton Conf. Communication, Control and Computing*, Monticello, IL, USA, September 2006.

[7] D. Hughes-Hartogs. The capacity of the degraded spectral Gaussian broadcast channel. Ph.D. dissertation, Stanford University, 1975.

[8] D. N. Tse. Optimal power allocation over parallel Gaussian broadcast channels. In *Proc. IEEE Int. Symp. Information Theory (ISIT)*, page 27, Ulm, Germany, June 1997.

[9] L. Li and A. J. Goldsmith. Capacity and optimal resource allocation for fading broadcast channels-Part I: ergodic capacity. *IEEE Trans. Inform. Theory*, 47(3):1083–1102, March 2001.

[10] L. Li and A. J. Goldsmith. Capacity and optimal resource allocation for fading broadcast channels-Part II: outage capacity. *IEEE Trans. Inform. Theory*, 47(3):1103–1127, March 2001.

[11] N. Jindal and A. Goldsmith. Optimal power allocation for parallel Gaussian broadcast channels with independent and common information. In *Proc. IEEE Int. Symp. Information Theory (ISIT)*, Chicago, Illinois, USA, June-July 2004.

[12] Y. Liang and H. V. Poor. Secure communication over fading channels. In *Proc. Annu. Allerton Conf. Communication, Control and Computing*, Monticello, IL, USA, September 2006.

[13] Z. Li, R. Yates, and W. Trappe. Secrecy capacity of independent parallel channels. In *Proc. Annu. Allerton Conf. Communication, Control and Computing*, Monticello, IL, USA, September 2006.

[14] P. Gopala, L. Lai, and H. El Gamal. On the secrecy capacity of fading channels. Submitted to *IEEE Trans. Inform. Theory*, Oct. 2006.

[15] H. Yamamoto. A coding theorem for secret sharing communication systems with two Gaussian wiretap channels. *IEEE Trans. Inform. Theory*, 37(3):634–638, May 1991.

[16] G. Caire, G. Taricco, and E. Biglieri. Optimal power control over fading channels. *IEEE Trans. Inform. Theory*, 45(5):1468–1489, July 1999.

[17] A. Carleial and M. Hellman. A note on Wyner's wiretap channel. *IEEE Trans. Inform. Theory*, 23(3):387–390, May 1977.

[18] S. K. Leung-Yan-Cheong. On a special class of wire-tap channels. *IEEE Trans. Inform. Theory*, 23(5):625–627, September 1977.

[19] H. Yamamoto. Coding theorem for secret sharing communication systems with two noisy channels. *IEEE Trans. Inform. Theory*, 35(3):572–578, May 1989.

[20] M. van Dijk. On a special class of broadcast channels with confidential messages. *IEEE Trans. Inform. Theory*, 43(2):712–714, March 1997.

[21] C. Mitrpant, Y. Luo, and A. J. H. Vinck. Achieving the perfect secrecy for the Gaussian wiretap channel with side information. In *Proc. IEEE Int. Symp. Information Theory (ISIT)*, page 44, Chicago, Illinois, June-July 2004.

[22] H. Koga and N. Sato. On an upper bound of the secrecy capacity for a general wiretap channel. In *Proc. IEEE Int. Symp. Information Theory (ISIT)*, pages 1641–1645, Adelaide, Australia, September 2005.

[23] G. Cohen and G. Zemor. The wire-tap channel applied to biometrics. In *Proc. Inter. Symp. Inform. Theory and its App. (ISITA)*, Parma, Italy, October 2004.

[24] A. Thangaraj, S. Dihidar, A. Calderbank, S. McLaughlin, and J.-M. Merolla. On achieving capacity on the wire tap channel using LDPC codes. In *Proc. IEEE Int. Symp. Information Theory (ISIT)*, Adelaide, Australia, September 2005.

[25] M. Hayashi. General nonasymptotic and asymptotic formulas in channel resolvability and identification capacity and their application to the wiretap channel. *IEEE Trans. Inform. Theory*, 52(4):1562–1575, April 2006.

[26] A. Khisti, G. Wornell, and A. Tchamkerten. Secure broadcasting with multiuser diversity. In *Proc. Annu. Allerton Conf. Communication, Control and Computing*, Monticello, IL, USA, September 2006.

[27] M. Bloch, J. Barros, M. Rodrigues, and S. W. McLaughlin. An opportunistic physical-layer approach to secure wireless communications. In *Proc. Annu. Allerton Conf. Communication, Control and Computing*, Monticello, IL, USA, September 2006.

[28] R. Ahlswede and I. Csiszár. Common randomness in information theory and cryptography-Part I: Secret sharing. *IEEE Trans. Inform. Theory*, 39(4):1121–1132, 1993.

[29] R. Ahlswede and I. Csiszár. Common randomness in information theory and cryptography-Part II: CR capacity. *IEEE Trans. Inform. Theory*, 44(1):225–240, January 1998.

[30] U. M. Maurer. Secrete key agreement by public discussion based on common information. *IEEE Trans. Inform. Theory*, 39(5):733–742, May 1993.

[31] I. Csiszár and P. Narayan. Common randomness and secret key generation with a helper. *IEEE Trans. Inform. Theory*, 46(2):344–366, March 2000.

[32] I. Csiszár and P. Narayan. Secrecy capacities for mulitple terminals. *IEEE Trans. Inform. Theory*, 50(12):3047–3061, December 2004.

[33] I. Csiszár and P. Narayan. Secrecy capacities for multiterminal channel models. In *Proc. IEEE Int. Symp. Information Theory (ISIT)*, pages 2138–2141, Adelaide, Australia, 2005.

[34] U. M. Maurer and S. Wolf. From weak to strong information-theoretic key agreement. In *Proc. IEEE Int. Symp. Information Theory (ISIT)*, page 18, Sorrento, Italy, June 2000.

[35] U. M. Maurer and S. Wolf. Secret-key agreement over unauthenticated public channels-Part I. definitions and a completeness result. *IEEE Trans. Inform. Theory*, 49(4):822–831, April 2003.

[36] U. M. Maurer and S. Wolf. Secret-key agreement over unauthenticated public channels-Part II. privacy amplification. *IEEE Trans. Inform. Theory*, 49(4):832–838, April 2003.

[37] U. M. Maurer and S. Wolf. Secret-key agreement over unauthenticated public channels-Part III. privacy amplification. *IEEE Trans. Inform. Theory*, 49(4):839–851, April 2003.

[38] S. Venkatesan and V. Anantharam. The common randomness capacity of a pair of independent discrete memoryless channels. *IEEE Trans. Inform. Theory*, 44(1):215–224, January 1998.

[39] S. Venkatesan and V. Anantharam. The common randomness capacity of a network of discrete memoryless channels. *IEEE Trans. Inform. Theory*, 46(2):367–387, March 2000.

[40] Y. Liang and H. V. Poor. Generalized multiple access channels with confidential messages. Submitted to *IEEE Trans. Inform. Theory*, April 2006; available at http://www.arxiv.org/PS_cache/cs/pdf/0605/0605014.pdf.

[41] R. Liu, I. Maric, R. Yates, and P. Spasojevic. The discrete memoryless multiple access channel with confidential messages. In *Proc. IEEE Int. Symp. Information Theory (ISIT)*, Seattle, Washington, USA, July 2006.

[42] E. Tekin and A. Yener. The Gaussian multiple access wire-tap channel. Submitted to *IEEE Trans. Inform. Theory*, April 2006.

[43] E. Tekin and A. Yener. Achievable rates for the general Gaussian multiple access wire-tap channel with collective secrecy. In *Proc. Annu. Allerton Conf. Communication, Control and Computing*, Monticello, IL, USA, September 2006.

[44] Y. Oohama. Coding for relay channels with confidential messages. In *Proc. IEEE Information Theory Workshop (ITW)*, pages 87–89, Cairns, Australia, September 2001.

[45] A. A. El Gamal. Capacity of the product and sum of two unmatched broadcast channels. *Probl. Inform. Transm.*, 16(1):1–16, Jan.-Mar. 1980.

[46] T. M. Cover and J. A. Thomas. *Elements of Information Theory.* Wiley, New York, 1991.

[47] Y. Liang, V. V. Veeravalli, and H. V. Poor. Resource allocation for wireless fading relay channels: max-min solution. Submitted to *IEEE Trans. Inform. Theory*, Aug. 2006.